\documentclass[a4paper,11pt]{article}
\usepackage{jcappub} 
\usepackage{lineno}

\usepackage{graphicx}
\usepackage{float}
\usepackage{amsmath}
\usepackage{amssymb}
\usepackage{url}
\usepackage{color}
\usepackage{xcolor}

\begin{document}

\title{\boldmath Cosmological Signatures of the Conformal and Non-Conformal Next-to-Minimal Two-Higgs-Doublet Model}

\author[a,1]{Filip Gustavsson \note{Corresponding author.}}
\author[a]{Roman Pasechnik}
\author[a]{Johan Rathsman}
 \affiliation[a]{\small\it Department of Physics, Lund University,\\221 00 Lund, Sweden}
 
\emailAdd{fi8225gu-s@student.lu.se}
\emailAdd{roman.pasechnik@fysik.lu.se}
\emailAdd{johan.rathsman@fysik.lu.se}

\abstract{We investigate cosmological signatures of the next-to-minimal two-Higgs-doublet model (N2HDM) in both the classically scale-invariant, or conformal, realization and in the corresponding non-conformal theory. Using a one-loop finite-temperature effective potential, collider and flavour constraints implemented through \texttt{ScannerS}, \texttt{HiggsBounds} and \texttt{HiggsSignals}, and a modified on-shell counterterm prescription for the conformal model, we identify parameter regions with a first-order electroweak phase transition and potentially observable stochastic gravitational waves. We find that both conformal and non-conformal scenarios can produce sound-wave signals in the sensitivity range of future space-based interferometers, including LISA, while the electroweak transitions complete promptly ($T_n\simeq T_p$), with pronounced supercooling below the critical temperature confined to the strongest transitions. The two scenarios nevertheless populate different regions of the scalar mass spectrum when the gravitational-wave signal is observable, suggesting that a future stochastic background measurement combined with collider information on the additional Higgs states could discriminate between the two realizations. We further study partonic Higgs-pair production, in the heavy-top approximation, and find that large deviations in the SM-like di-Higgs rate tend to occur away from the regions with the strongest gravitational-wave signals, while enhanced BSM Higgs-pair rates are mainly associated with the conformal model. The analysis highlights a complementary interplay between electroweak cosmology, scalar spectroscopy and Higgs self-interaction probes in extended Higgs sectors.}

\maketitle

\flushbottom

\section{Introduction}\label{sec::Introduction}
In 2012 the Higgs boson was discovered \cite{ATLAS:Discovery2012,CMS:Discovery}, completing the set of elementary particles predicted by the Standard Model (SM). A natural continuation is then to study what type of beyond the Standard Model (BSM) physics is possible in the Higgs sector, including the ways electroweak symmetry can be broken as  characterized by the electroweak phase transition (EWPT). This possibility still needs to be investigated, since the current experimental uncertainties leave enough room for BSM contributions \cite{ATLAS:Nature,CMS:Nature}, particularly in the Higgs sector. These extensions can add important BSM physics, such as a strong first-order electroweak phase-transition, which can satisfy one of the conditions needed for baryogenesis during the electroweak transition. 

Baryogenesis can occur in the early universe during the EWPT. This occurs when a field configuration changes from being centered around one minimum of the scalar potential to another. During the phase transition baryogenesis can occur if certain conditions, the Sakharov conditions \cite{Sakharov:Conditions}, are met. These are: existence of processes that violate baryon number, departure from thermal equilibrium during these processes as well as C and CP-violation. The SM undergoes a crossover phase transition \cite{Kajantie:PT,DOnofrio:2012phz}, therefore, the out of thermal equilibrium condition fails in addition to the amount of CP-violation in the SM being too small to explain the observed matter-antimatter asymmetry \cite{Huet::CPSM}. Finding out which BSM models feature a first-order transition strong enough to satisfy the Sakharov conditions remains an open area of research \cite{Athron:PhaseTracer2,Dorsch:2HDMEWPT,Ghosh:CxSMEWPT}. 

During the EWPT gravitational waves (GWs) can be created, referred to as primordial GWs. After the first observation of GWs \cite{LIGO:GWDiscovery} it is natural, as a continuation, to search for primordial GWs. There are many planned future experiments that have this as a part of their goal, such as the LISA experiment which hopes to measure, or provide a bound, on primordial GWs. Therefore, it is reasonable to investigate which BSM models have first-order transitions that release observable GWs. This has been studied for several models; examples include the Two-Higgs-Doublet Model (2HDM) \cite{Caprini:LISADetetability}, a model with a scalar triplet \cite{Friedrich:GWTriplet} and the SM with a real scalar \cite{Braathen:RxSMGW}. 

In this paper we instead analyze the Next-to-minimal Two-Higgs-Doublet Model (N2HDM) and study the possibility of GW detection at LISA and what conclusions could be drawn from such an observation. Similarly to the 2HDM, the N2HDM adds a Higgs doublet to the SM, as well as an additional singlet. 
In addition to analysing the general N2HDM, we also study its scale invariant version, where there are no mass terms present in the potential that has a classical scale-invariance (SI). The SI gives the model unique properties such as a scalar particle which is massless at tree-level attaining mass through radiative corrections. In addition to the SI, a $\mathbb Z_2$-symmetry is imposed to disallow tree-level flavour-changing neutral currents (FCNCs) as these are tightly constrained \cite{ATLASFCNC}.

The collider phenomenology of the N2HDM has been studied extensively \cite{Arhrib:N2HDMPheno,Paasch:THDMSThesis,Muhlleitner:N2HDM}, and its electroweak phase-transition dynamics has also been explored in recent work \cite{Biermann:N2HDMEWPT,Basler:nonMinimalEWPT,Chaudhuri:2024N2HDMGW}. 
The purpose of the present study is to make a systematic comparison between the ordinary, non-SI N2HDM and its classically SI counterpart under a common set of theoretical, collider and flavour constraints. In particular, we ask whether a stochastic GW signal from a first-order EWPT, together with information on the scalar spectrum and Higgs-pair production, could discriminate between a conformal and non-conformal N2HDM. Our analysis is therefore not primarily a demonstration that the N2HDM can support a first-order electroweak transition that fulfills all Sakharov conditions; rather, its main purpose is to determine whether the classically SI and non-SI realizations populate distinguishable regions of the parameter space when considering both gravitational waves and collider data.

The paper is organized as follows. Section~\ref{sec::Model} defines the N2HDM, its SI limit and the one-loop renormalization prescription used in the conformal and non-conformal case. Section~\ref{sec::Thermal} summarizes the finite-temperature effective potential, the phase transition observables and the GW treatment. Section~\ref{sec::Scan} describes the theoretical and phenomenological constraints and the parameter scan. The EWPT and GW results are presented in Sec.~\ref{sec::EWPT}, while Sec.~\ref{sec::DiHiggs} studies the complementarity of partonic Higgs-pair production and GW production. Lastly, we draw conclusions in Sec.~\ref{sec::Conclusion}.

\section{The Model}\label{sec::Model}
The N2HDM extends the 2HDM with a scalar singlet such that the model can be parametrized using two doublets, $\Phi_i$, and a real singlet, $S$:
\begin{equation}
    \Phi_1=\begin{pmatrix}
        \rho_1^+\\
        \frac{1}{\sqrt 2}(v_1+\varphi_1+i\eta_1)
    \end{pmatrix},\quad
    \Phi_2=\begin{pmatrix}
        \rho_2^+\\
        \frac{1}{\sqrt 2}(v_2+\varphi_2+i\eta _2)
    \end{pmatrix},\quad
    S=v_s+\varphi_s. \label{eq::BasisDef}
\end{equation}
Here $\varphi_i$ denote the CP-even neutral fields, $\eta_i$ the CP-odd neutral fields, $\rho_i^\pm$ the charged fields and $v_i$ the vacuum expectation values (VEVs). We work in the generic basis in which the scalar fields transform under the imposed $\mathbb Z_2$ symmetries:
\begin{align}
    \mathbb Z_2:\quad  \Phi_1\to \Phi_1,\quad \Phi_2\to -\Phi_2,\quad S\to S,\\
    \mathbb Z_2': \Phi_1\to \Phi_1,\quad \Phi_2\to \Phi_2,\quad S\to -S.
\end{align}
By assigning appropriate $\mathbb Z_2$ symmetries to the right-handed fermion fields it is possible to suppress tree-level flavour-changing neutral currents \cite{FCNC} in the Yukawa couplings, which are tightly constrained experimentally \cite{ATLASFCNC}. In the scalar potential the doublet $\mathbb{Z}_2$-symmetry enforces natural flavour conservation, forbidding tree-level flavour-changing neutral currents, and is softly broken by the $m_{12}^2$ term.  The second symmetry, $\mathbb{Z}_2'$, forbids odd powers of the singlet field and is spontaneously broken once $S$ acquires a vacuum expectation value $v_s\neq0$\footnote{Since $Z_2'$ is an exact symmetry of the potential that is spontaneously broken by $v_s$, its breaking would seed a network of cosmological domain walls. We assume a small explicit $Z_2'$-breaking term, phenomenologically negligible for the observables studied here, which lifts the degeneracy between the $\pm v_s$ vacua so that the walls collapse before dominating the energy density.}; this breaking is the origin of the singlet--doublet mixing quantified by $\alpha_2$ defined below, with $\sin{\alpha_2}=v_s/\sqrt{v_1^2+v_2^2+v_s^2}$.

The corresponding CP-conserving tree-level potential, with the soft doublet-breaking term included, is then given by
\begin{align}
V_0(\Phi_1,\Phi_2,S)
&= m_{11}^2(\Phi_1^\dagger\Phi_1)
 + m_{22}^2(\Phi_2^\dagger\Phi_2)
 - m_{12}^2(\Phi_1^\dagger\Phi_2+\Phi_2^\dagger\Phi_1) +\frac{m_S^2}{2}S^2             \notag\\[4pt]
&+\frac{\lambda_1}{2}(\Phi_1^\dagger\Phi_1)^2
 +\frac{\lambda_2}{2}(\Phi_2^\dagger\Phi_2)^2
 +\lambda_3(\Phi_1^\dagger\Phi_1)(\Phi_2^\dagger\Phi_2)
 +\lambda_4(\Phi_1^\dagger\Phi_2)(\Phi_2^\dagger\Phi_1)               \label{eq:PotGeneric}  \\[4pt]
& +\frac{\lambda_5}{2}\!\left[(\Phi_1^\dagger\Phi_2)^2
   +(\Phi_2^\dagger\Phi_1)^2\right]
 +\frac{\lambda_6}{8}S^4
 +\frac{S^2}{2}\!\left[\lambda_7(\Phi_1^\dagger\Phi_1)
                       +\lambda_8(\Phi_2^\dagger\Phi_2)\right].     \notag
\end{align}
Here $\lambda_i$ are real quartic couplings and $m_{ij}^2$ are quadratic mass parameters. Taking the scalar potential to be CP conserving avoids CP-even/CP-odd mixing in the neutral sector. Consequently, the present work does not attempt to realize electroweak baryogenesis in full; it rather focuses on the out-of-equilibrium Sakharov condition associated with a first-order EWPT.

In the classically SI limit all dimensionful parameters in Eq.~\eqref{eq:PotGeneric} are set to zero, $m_{11}^2=m_{22}^2=m_{12}^2=m_S^2=0$. The tree-level potential is then quartic and has a flat direction given by the Gildener--Weinberg conditions \cite{Gildener:Conformal}. Electroweak symmetry breaking is generated radiatively once the SI is broken by the one-loop effective potential. In the finite-temperature discussion below we refer to this classically SI model as the conformal N2HDM, while keeping in mind that classical SI and full quantum conformal invariance are not identical concepts.

From the potential, the tadpole conditions can be defined as the equations fulfilled when the tree-level potential has an extremum at the VEV. It follows that: 

\begin{align}
    \dfrac{\partial V_0}{\partial \varphi_1}\bigg|_{\Phi=\langle\Phi\rangle,S=\langle S\rangle}&= m_{11}^2v_1-m_{12}^2v_2+\frac{\lambda_1}{2}v_1^3+\frac{\lambda_{345}}{2}v_1v_2^2+\frac{\lambda_7}{2}v_1v_s^2=0,\label{eq::tadpole1}\\
    \dfrac{\partial V_0}{\partial \varphi_2}\bigg|_{\Phi=\langle\Phi\rangle,S=\langle S\rangle}&= m_{22}^2v_2-m_{12}^2v_1+\frac{\lambda_2}{2}v_2^3+\frac{\lambda_{345}}{2}v_1^2v_2+\frac{\lambda_8}{2}v_2v_s^2=0,\label{eq::tadpole2}\\
    \dfrac{\partial V_0}{\partial \varphi_s}\bigg|_{\Phi=\langle\Phi\rangle,S=\langle S\rangle}&= m_S^2v_s+\frac{\lambda_6}{2}v_s^3+ \frac{v_s}{2}(\lambda_7v_1^2+\lambda_8v_2^2)=0.\label{eq::tadpole3}
\end{align}
where $\lambda_{345}=\lambda_3+\lambda_4+\lambda_5$. These three conditions are used to eliminate three parameters in the potential in favor of the VEVs. In the SI case the tree-level potential is flat along the vacuum direction; the physical electroweak scale is then fixed only after including the radiative corrections that lift this flatness.

A transformed basis for the doublets can be defined such that the VEVs of the doublets appear in only one of the doublets, $H_1$. This results in the \textit{Higgs basis}: 
 \begin{equation}
    H_1=\begin{pmatrix}
        G^+\\
        \frac{1}{\sqrt 2}(v+h'_1+iG_0)
    \end{pmatrix},\quad
    H_2=\begin{pmatrix}
        H^+\\
        \frac{1}{\sqrt 2}(h'_2+iA)
    \end{pmatrix}.
\end{equation}
This basis is used when diagonalizing the mass matrix at tree-level. The transformation between the bases is defined using a rotation of the doublets by the angle $\beta$: 
\begin{equation}
    t_\beta=\frac{v_2}{v_1}, \quad v_1^2+v_2^2=v^2,\label{eq::BetaDefinition}
\end{equation}
where the abbreviations 
\begin{equation}
    t_\beta=\tan\beta, \quad s_\beta=\sin\beta, \quad c_\beta=\cos\beta, 
\end{equation}
are used for the trigonometric functions.  The relation between the Higgs basis and the generic basis can now be written as 
 \begin{equation}
     \begin{pmatrix}
        H_1\\H_2
    \end{pmatrix}
    =\begin{pmatrix}
        c_\beta& s_\beta\\
        -s_\beta&c_\beta
    \end{pmatrix}
    \begin{pmatrix}
        \Phi_1\\ \Phi_2
    \end{pmatrix}.\label{eq::BasisRot}
 \end{equation}

Taking the real and imaginary components of the scalar fields as a basis, in vector form: $\vec \phi$, the neutral and charged components form a $9\times9$ real symmetric field-dependent mass matrix, which is block diagonal in the CP-conserving vacuum. The matrix is split into two 2$\times$2 blocks corresponding to charged particles, one 2$\times$2 block corresponding to CP-odd particles and one 3$\times$3 block corresponding to the CP-even particles. The problem of diagonalizing the mass matrix is therefore equivalent to diagonalizing each of these blocks. 

The tree-level mass matrix is given by 
\begin{equation}
    \mathcal{M}^{(0)}_{ij}=\dfrac{\partial^2V_0(\vec \phi)}{\partial\phi_i\partial\phi_j}\bigg|_{\vec 
    \phi=0},\label{eq::MassMatrixDef}
\end{equation}
which, in the mass eigenstates, written in the basis previously defined, is diagonalized as
\begin{equation}
    \mathcal{M}^{(0)}_{diag}=\text{diag}(m_{h_1}^2,m_{h_2}^2,m_{h_3}^2,0,m_A^2,0,m_{H^\pm}^2,0,m_{H^\pm}^2).
\end{equation}
Switching to the Higgs basis will diagonalize the CP-odd and charged blocks of the mass matrix giving
\begin{align}
    m_{H^\pm}^2&=-\frac{1}{2}\left(\lambda_4+\lambda_5\right)v^2+\frac{m_{12}^2}{c_\beta s_\beta} \, ,\\
    m_A^2&=-\lambda_5v^2+\frac{m_{12}^2}{c_\beta s_\beta} \, . \label{eq::mA}
\end{align}
The other entries in the diagonalized blocks of the CP-odd and charged particles will be zero as they are associated with the Goldstone bosons. 

Defining a diagonalized form of the remaining CP-even matrix, $\mathcal{M}_{diag}$, one can relate it to the original mass matrix. This is done through three rotations, parameterized by $\alpha_1$, $\alpha_2$ and $\alpha_3$:
\begin{equation}
    \mathcal{M}_{gen}=R_1^\dagger R_2^\dagger R_3^\dagger\mathcal{M}_{diag}R_3 R_2R_1,
\end{equation}
where the rotation matrices are defined as
\begin{align}
     R_1&=\begin{pmatrix}
        c_{\alpha_1}&-s_{\alpha_1}&0\\
        s_{\alpha_1}&c_{\alpha_1}&0\\
        0&0&1
    \end{pmatrix},\quad
    R_2&=\begin{pmatrix}
        c_{\alpha_2}&0&-s_{\alpha_2}\\
        0&1&0\\
        s_{\alpha_2}&0&c_{\alpha_2}
    \end{pmatrix},\quad
    R_3&=\begin{pmatrix}
        1&0&0\\
        0& c_{\alpha_3} & -s_{\alpha_3}\\
        0&s_{\alpha_3}&c_{\alpha_3}
    \end{pmatrix}. \label{eq::RotationMatrix}
 \end{align}
In the case where one of the particles is a SM-like Higgs we refer to it as $h$ whereas the other CP-even Higgses are denoted as $l$ and $H$ with the convention
\begin{equation}
    m_l<m_H.
\end{equation}
We can now write $R=R_1R_2R_3$ and denote
\begin{align}
    &\begin{pmatrix}
        h_1\\
        h_2\\
        h_3
    \end{pmatrix}=R\begin{pmatrix}
        \varphi_1\\
        \varphi_2\\
        \varphi_s
    \end{pmatrix},\\
&R=\begin{pmatrix}
c_{\alpha_1}c_{\alpha_2} & -c_{\alpha_3}s_{\alpha_1}-c_{\alpha_1}s_{\alpha_2}s_{\alpha_3} & s_{\alpha_1}s_{\alpha_3}-c_{\alpha_1}c_{\alpha_3}s_{\alpha_2}\\
c_{\alpha_2}s_{\alpha_1} & c_{\alpha_1}c_{\alpha_3}-s_{\alpha_1}s_{\alpha_2}s_{\alpha_3} & -c_{\alpha_1}s_{\alpha_3}-c_{\alpha_3}s_{\alpha_1}s_{\alpha_2}\\
s_{\alpha_2} & c_{\alpha_2}s_{\alpha_3} & c_{\alpha_2}c_{\alpha_3}
\end{pmatrix}.\label{eq::RMatrix}
\end{align} 
where $h_i$ are the mass eigenstates.

 In the SI N2HDM, with the VEV being in the flat direction, one scalar particle, $h_1$, will be directed along this direction and have $m_{h_1}=0$. Knowing the direction of one of the Higgs particles, it follows that two of the rotation angles are given as ratios of VEVs.  
 In a model with  $\mathbb Z_2$-symmetries, this gives
\begin{align}
    \alpha_1&=\pm \beta, \label{eq::alpha1}\\
    s_{\alpha_2}&=\pm\frac{v_s}{v_h}, \label{eq::alpha2}
\end{align}
where $v_h$ is the total VEV: 
\begin{equation}
     v_h=\sqrt{v^2+v_s^2}. \label{eq::vhDefinition}
\end{equation}
As already mentioned, another consequence of this flat direction is that $h_1$ is massless at tree-level. Following Gildener and Weinberg~\cite{Gildener:Conformal} we refer to it as the \textit{scalon}.

In addition to the scalar self-interactions, the Higgs particles will also interact with fermions and gauge bosons. The interactions with the gauge bosons will give a constraint on the rotation angles defined in Eq.~\ref{eq::RotationMatrix}. This 
follows from the $h_iVV$-couplings, where $V$ is a vector boson, which is known to constrain the 2HDM \cite{ATLAS:2HDMConstraints}. These tree-level couplings are given by 
\begin{equation}
    \frac{c(h_iVV)}{c(hVV)_{\rm SM}}= c_{\beta}R_{i1}+s_{\beta}R_{2i}.
\end{equation}

In the N2HDM this means the allowed parameter space will be near to the alignment limit, which is defined as the limit where the SM-like Higgs has the same coupling to the gauge bosons as in the SM. From inspection of the couplings, given in Tab.~\ref{tab:eff_couplings}, alignment can be reached in two different scenarios: 
\begin{align}
    \text{A:}& \quad \alpha_1\to \beta , \, \alpha_2\to 0, \quad (h_1 \, \text{ SM-like}), \label{eq::scenarioA} \\
    \text{B:}& \quad \alpha_1\to \beta , \, \alpha_2\to \pm \frac{\pi}{2}, \, \alpha_3\to 0, \quad (h_2  \,  \text{ SM-like}).
\end{align}

\begin{table}
\centering
\caption{The effective couplings of the CP-even Higgs particles in the N2HDM, $h_i$, to the gauge bosons in relation to the SM-values. Where $R$ is the matrix defined by Eq. \ref{eq::RMatrix}.}
\vspace*{0.2cm}
\begin{tabular}{|c|c|c|}
\hline
 $c(h_1VV)/c(hVV)_{\rm SM}$ & {$c(h_2VV)/c(hVV)_{\rm SM}$}& $c(h_3VV)/c(hVV)_{\rm SM}$ \\
\hline
$c_{\beta-\alpha_1}c_{\alpha_2}$ &
$-s_{\beta-\alpha_1}s_{\alpha_2}s_{\alpha_3} -c_{\beta-\alpha_1} c_{\alpha_3}$ &
$-s_{\beta-\alpha_1}s_{\alpha_2}c_{\alpha_3} +c_{\beta-\alpha_1}s_{\alpha_3}$ \\
\hline
\end{tabular}
\label{tab:eff_couplings}
\end{table}

For the Type-I Yukawa realization all fermions couple to the same doublet, $\Phi_2$. Let $Q_L$ refer to the weak isospin doublet containing quarks, $u_R$ to up-type right-handed quarks and $d_R$ to down-type right-handed quarks. Similarly, let $L_L$ refer to the weak isospin doublet containing leptons and $\ell_R$ to right-handed charged leptons. In a gauge-invariant notation the Yukawa sector can then be written as
\begin{equation}
    -\mathcal{L}_Y = \bar Q_L^i (Y_u)_{ij}\tilde\Phi_2 u_R^j + \bar Q_L^i (Y_d)_{ij}\Phi_2 d_R^j + \bar L_L^i (Y_\ell)_{ij}\Phi_2 \ell_R^j + \mathrm{h.c.}\label{eq::YukawaLag}
\end{equation}
Here $i,j$ labels fermion generations. Due to the $\mathbb Z_2$-symmetry, the Yukawa couplings are diagonal in the fermion mass state basis. The corresponding normalized top-Yukawa couplings entering the gluon-fusion amplitudes are obtained by rotating the CP-even interaction eigenstates to the mass basis with the matrix in Eq.~\eqref{eq::RotationMatrix}.

In addition to the tree-level potential, there are one-loop level effects which enter the effective potential. At zero temperature, the effective potential is split into 
\begin{equation}
    V_{eff}=V_0+V_{CW}+V_{CT},
\end{equation}
where $V_0$ is the tree-level potential and the other terms are one-loop contributions. $V_{CT}$ is the contribution from the finite part of the counterterms (CTs) which are introduced during the renormalization procedure while $V_{CW}$ is the Coleman-Weinberg (CW) potential and corresponds to proper one-loop contributions calculated in the $\overline{\textrm{MS}}$ scheme. The CW potential is well known and has the form \cite{CW:Thesis}
\begin{equation}
    V_{CW}(\vec \phi)=\frac{1}{4(4\pi)^2}\sum_{X=S,G,F}(-1)^{2s_X}n_X\text{Tr}\left[\mathcal{M}^4_{(X)}\left(\log\left(\frac{\mathcal{M}^2_{(X)}}{\mu^2}\right)-k_X\right)\right],\label{eq::CW}
\end{equation}
where $X=S,G,F$ denote sums over different types of particles: scalars (S), gauge bosons (G) and fermions (F), while $n_X$ is the associated number of degrees of freedom for a particle of type $X$. The constants that enter the equation are: $\mu$, the renormalization scale, which we set to $v_h$, and $k_X$, which are given by 
\begin{equation}
    k_X=\begin{cases}
        \frac{5}{6}\quad X=G,\\
        \frac{3}{2}\quad X=S,F.
    \end{cases}
\end{equation}
The tree-level mass matrices that enter the expression, $\mathcal{M}^2_{(X)}$, are field-dependent and the scalar part is given by:
\begin{equation}
   \left(\mathcal{M}^2_{S}\right)_{ij}=\dfrac{\partial^2 V_{0}}{\partial \phi_i\partial \phi_j}.
\end{equation}
In other words, $ \mathcal{M}^2_{S}$ is not evaluated at the VEV,  instead the field dependence is kept which is how the fields enter Eq. \eqref{eq::CW}.

The gauge bosons and fermions provide a contribution to the CW potential coming from their interactions with the scalars and their mass matrices are 
\begin{align}
    \left(\mathcal{M}_{F}\right)_{ij}&=-\dfrac{\partial\mathcal{L}_{Y}}{\partial \bar \psi_i\partial \psi_j},\\
    \left(\mathcal{M}^2_{G}\right)_{ij}&=-\dfrac{\partial\mathcal{L}_{G}}{\partial A_{i \, \mu}\partial A^\mu_j},
\end{align}
where $A^\mu$ is a gauge boson and $\psi$ is a fermionic field. Using these matrices the CW potential is defined and can be calculated.

The final contribution to the effective potential, $V_{CT}$, is the contribution to the scalar potential that arises from the finite parts of the CTs, which appear when $\overline{\textrm{MS}}$ is not used. Here we use an on-shell (OS) scheme with the renormalization condition introduced later in this section. 

For the CT-potential we use a potential of the same form as Ref.~\cite{Basler:BSMPT1},
\begin{equation}
\begin{split}
    V_{CT}&=V_0(m\to \delta m,\lambda\to \delta \lambda)+\delta T_1(v_1+\varphi_1)+\delta T_2(v_2+\varphi_2)+\delta T_S(v_s+\varphi_S),
\end{split}\label{eq::VCT_corrected}
\end{equation}
where the tadpole CTs are associated with the respective CP-even fields. We thus assume that there are no tadpole CTs for the CP-odd or charged fields\footnote{The notation $V_0(m\to \delta m,\lambda\to \delta \lambda)$ denotes the tree-level potential in Eq.~\eqref{eq:PotGeneric} with all quadratic and quartic parameters replaced by the corresponding CT coefficients.}.

In order to determine the values of the CT parameters we have to impose the renormalization conditions. Here we follow Ref~\cite{Basler:BSMPT1} and use the OS renormalization scheme, which ensures the masses of the scalar particles and the location of the minima of the potential are unchanged by higher-order corrections. 
This means that the CT parameters can be derived from 
\begin{align}
    \partial_{\phi_i}V_{CW}\bigg|_{\vec \phi =0}&=-\partial_{\phi_i}V_{CT}\bigg|_{\vec \phi =0},\\
    \partial_{\phi_i}\partial_{\phi_j}V_{CW}\bigg|_{\vec \phi =0}&=-\partial_{\phi_i}\partial_{\phi_j}V_{CT}\bigg|_{\vec \phi =0}.\label{eq::SecondOrderCondtion}
\end{align}
To see the explicit formulas that this leads to see Ref.~\cite{Basler:BSMPT1}. 

In the SI case this scheme needs to be modified to take into account the flat direction in the tree-level potential. If the OS scheme is used as defined above, the scalon would remain massless even at higher orders, where the flat-direction is lifted. This follows since the CT conditions imposed on the effective potential enforce the scalon to not change its mass when going to higher orders, even though the SI is broken. Therefore, the OS scheme needs to be modified to allow for the scalar potential to have a symmetry at tree-level which is broken through radiative corrections. This modification, inspired by Ref.~\cite{Rustas:Thesis}, enters into Eq. \eqref{eq::SecondOrderCondtion} as a shift to the mass of the scalon such that
\begin{equation}
    \partial_{\phi_i}\partial_{\phi_j}V_{CW}\bigg|_{\vec \phi =0}+\partial_{\phi_i}\partial_{\phi_j}V_{CT}\bigg|_{\vec \phi =0}=R ^\dagger\text{diag}(m_{h_1}^2,0,0,...)R ,\label{eq::ModifiedOS}
\end{equation}
where $R$ is the rotation matrix derived for the CP-even masses at tree-level and $m_{h_1}$ is the one-loop correction to the mass of the scalon. This gives additional contributions to the CT parameters, which are provided in appendix \ref{sec::AppendixCT}. We emphasize that this modification of the scheme is only used for the SI case while the original OS scheme is used for the non-SI case.

\section{Thermal Effects and Gravitational Waves}\label{sec::Thermal}
The effective potential, as so far described, is only valid at zero temperature, therefore, additional contributions are needed for temperature-dependent effects. As a consequence, the classical background fields will no longer be static VEVs but instead vary with temperature, $\vec \omega(T)$. This is a vector in the field configuration space and comparing this to the previous notation leads to
\begin{equation}
    \omega_i(T=0)=v_i.
\end{equation}
It then follows that the temperature-dependent effective potential is of the form
\begin{equation}
    V_{eff}(\vec \omega, T)=V_0(\vec \omega)+V_{CW}(\vec \omega)+V_{CT}(\vec\omega)+V_T(\vec\omega,T),
\end{equation}
where $V_T$ are the added thermal contributions. For details on how to calculate the thermal contribution at the one-loop level see e.g. Ref.~\cite{Quiros:Review}.

Our calculation of $V_{eff}$ uses the one-loop finite-temperature potential with daisy resummation as implemented in \texttt{BSMPT}~\cite{Basler:BSMPT1,Basler:BSMPT2,Basler:BSMPT3}. Only the bosonic Matsubara zero modes require resummation; in practice this amounts to replacing the field-dependent scalar and longitudinal gauge-boson masses by their thermally corrected Debye masses in the resummed part of the potential. The transverse gauge modes are not Debye screened. This convention must be kept fixed when comparing the conformal and non-conformal scans, since different resummation prescriptions, such as the Parwani \cite{Parwani:Resum} or Arnold--Espinosa schemes \cite{Arnold:Resum}, can shift the numerical values of observables. The plots below should therefore be read as results within a specified perturbative finite-temperature setup rather than as scheme-independent observables.

In the numerical analysis we use the same gauge, renormalization prescription and thermal-resummation setup for the conformal and non-conformal scans with the exception of the modification to the OS scheme in the conformal model previously introduced. The results therefore provide a controlled comparison between the two realizations, but the transition temperatures and derived quantities should not be viewed as strictly gauge-invariant observables at fixed perturbative order. The gauge and resummation dependence of the one-loop thermal effective potential should be regarded as a leading theoretical systematic uncertainty in the calculation of observables.

In practice, the potential is evaluated for the N2HDM using the program \texttt{BSMPT}~\cite{Basler:BSMPT1,Basler:BSMPT2,Basler:BSMPT3}, version/branch \texttt{3.0.7}. The modified renormalization conditions for the SI case, see Eq.~\ref{eq::ModifiedOS}, have been implemented in \texttt{BSMPT} such that these calculations are possible in the SI model. The finite-temperature potential is evaluated in Landau gauge, and daisy resummation is implemented using the \textit{Arnold-Espinosa} prescription. 
For the gravitational-wave spectra a fixed bubble-wall velocity $v_w=0.95$ is used. The quoted SNR values should therefore be interpreted within this common perturbative setup and wall-velocity assumption.

The temperature dependent potential allows for a change in phase, or in other words a transition from one minimum of the potential to another, and therefore for two phases to coexist. At the critical temperature, $T_c$, the minima corresponding to each phase will be equal such that 
\begin{equation}
    V_{eff}(\vec \omega =\vec \omega_1 ,T_c)=V_{eff}(\vec \omega =\vec \omega_2 ,T_c),
\end{equation}
where $\vec \omega_i$ represents the location of the two phases at the critical temperature. 

There are further temperatures that characterize the transition, the most important being 
the nucleation temperature $T_n$, defined as the temperature at which the number of critical bubbles nucleated per Hubble four-volume becomes of order unity. The definition of the nucleation temperature uses the Hubble parameter, which characterizes the expansion rate of the Universe, and the transition rate between the vacuua, $\Gamma$, the nucleation temperature is then given by:
\begin{equation}
    \frac{\Gamma}{H^4}(T_n)=1.
\end{equation}
This means that, at $T_n$, the rate of transition per Hubble volume is equal to the Hubble rate, $H$.

The percolation temperature, $T_p$, is defined by a large, connected structure of the true phase spanning the Universe. This is quantitatively defined as the temperature for which the new phase fills 29$\%$ of the Universe. Thus, the nucleation temperature is an indicator of when the rate of transition between phases is large while the percolation temperature indicates when the new phase occupies a significant volume of the Universe.

To supplement the temperatures presented, the phase transition can be additionally characterized by the strength, $\alpha$, and time-scale, $\frac{\beta}{H}$, of the phase transition. The parameter $\alpha$ is defined as the ratio of the energy density released by an expanding bubble of the new phase to the energy density of the surrounding photon gas \cite{Hindmarsh:2015qta,Hindmarsh:2017gnf}. This gives an estimate of the amount of energy which is available for GWs, as a larger $\alpha$ means more released energy. Meanwhile, $\beta^{-1}$ is the characteristic time scale of the PT and $H^{-1}$ is the characteristic time scale of the Universe at the time the PT occurs. Therefore, a large $\frac{\beta}{H}$ corresponds to a situation where the age of the Universe is much larger than the time scale of the PT while for smaller values they are of a similar magnitude. 

In the following we present the expressions for the parameters defined above that we use, for a derivation see the review Ref.~\cite{Athron:Review}. 
The parameter $\alpha$ is given by
\begin{equation}
    \alpha(T)=\frac{1}{\rho_{\rm rad}(T)}\left[\Delta V(T)-\frac{T}{4}\frac{d\Delta V(T)}{dT}\right],\qquad
    \Delta V(T)\equiv V(\vec\omega_f,T)-V(\vec\omega_t,T), \label{eq::alpha_corrected}
\end{equation}
where $\rho_{\rm rad}(T)=(\pi^2/30)g_*(T)T^4$ is the radiation energy density, $\vec\omega_t$ is the location of the true phase and $\vec\omega_f$ is the location of the false phase. 

Next we turn to $\frac{\beta}{H}$. This is defined using the 3 dimensional action given by
\begin{equation}
    S_3=4\pi \int d\rho \, \rho^2 \left[\frac{1}{2} \left( \dfrac{d \vec \phi}{d\rho}\right)^2+V(\vec\phi) \right] ,
\end{equation}
where $\rho=\sqrt{x_1^2+x_2^2+x_3^2}$ is the $O(3)$-symmetric radial coordinate and $\vec\phi$ denotes the scalar background-field connecting the false and true phases. Minimizing the action then yields a set of equations that are known as the bounce equations. The rate of transition between the phases can then be found as \cite{Callan:FalseVacuum}
\begin{equation}
    \Gamma=A(T)e^{-S_3/T}, \quad A(T)=T^4\left(\frac{S_3}{2\pi T}\right)^{3/2}.
\end{equation}
Using $dT/dt=-HT$ during radiation domination, the inverse duration is defined as
\begin{equation}
    \frac{\beta}{H}(T)=T\,\frac{d}{dT}\left(\frac{S_3}{T}\right), \label{eq::beta_corrected}
\end{equation}
where the derivative is evaluated at the relevant temperature.

Unless stated otherwise, the gravitational-wave quantities quoted below are evaluated at the percolation temperature, $T_\star=T_p$, as it provides a good estimate of the end of the EWPT. When comparing with results in the literature, quoted at the nucleation temperature, this convention should be kept in mind since $\alpha$ and $\beta/H$ can shift mildly between $T_n$ and $T_p$ even when $T_n/T_p-1$ is small.

A first-order transition can source a stochastic GW background through several mechanisms: scalar-field gradients during bubble collisions, long-lived acoustic waves in the plasma and magnetohydrodynamic turbulence. For electroweak-scale transitions in which the bubble walls do not run away, the sound-wave source is usually expected to dominate. The expanding bubbles inject kinetic energy into the surrounding plasma; the resulting bulk fluid motion sources tensor perturbations until the acoustic period ends, after which part of the energy may cascade into turbulence. We therefore use the sound-wave contribution as the baseline signal and treat bubble-collision and turbulence contributions as theoretical systematic uncertainties rather than as part of the central numerical prediction. For reviews and recent LISA-oriented treatments see Refs.~\cite{Athron:Review,Caprini:LISADetetability,Caprini:GWTheory}.

We have chosen to use the sound-wave contribution as the baseline source, since this is appropriate for non-runaway electroweak-scale transitions in which a sizable fraction of the released vacuum energy is transferred to the plasma. For very strong transitions, however, the finite lifetime of the acoustic source and possible turbulence or bubble-collision contributions can modify the amplitude. We therefore interpret the reported SNR values as estimates within the sound-wave approximation and explicitly flag very large-$\alpha$ values as points where a refined hydrodynamical treatment would be needed before making a precision forecast for LISA.

Unless stated otherwise, the quoted SNR values do not include an additional point-by-point acoustic-lifetime suppression factor beyond the baseline GW prescription used in the scan. For transitions with $\alpha\gtrsim1$, this can overestimate the sound-wave contribution, and a dedicated hydrodynamical calculation would be required to decide whether the bubble walls remain non-runaway and how much energy is transferred into sound waves, turbulence and scalar-field gradients \cite{Hindmarsh:2015qta,Hindmarsh:2017gnf,Ellis:2018mja,Caprini:GWTheory}.

The GWs formed in the EWPT will have certain key properties that can be observed. For instance, the power spectrum, $\Omega_{\rm GW}$, which shows how the GW energy varies with frequency, $f$, is defined as 
\begin{equation}
    \Omega_{\rm GW}=\frac{1}{\rho_c}\dfrac{\partial \rho_{\rm GW}}{\partial \ln(f)},
\end{equation}
where $\rho_c$ is the critical density of the Universe and $\rho_{GW}$ is the energy density of the GW. To characterize the power spectrum we use the peak frequency $f_{\rm peak}$ and the magnitude of the peak, $\Omega_{\rm peak}$, defined as
\begin{equation}
    \Omega_{\rm peak}=  \Omega_{\rm GW}(f=f_{\rm peak}).
\end{equation}
Introducing $h$ to characterize the present Hubble rate, $H_0$, as
\begin{equation}
    H_0=h\cdot 100\,\text{km}\,\text{s}^{-1}\,\text{Mpc}^{-1},
\end{equation}
the combination $h^2\Omega_{peak}$  can be used to estimate the GW spectra without explicit dependence on $H_0$. In addition to $\alpha$, the power spectrum also requires knowledge of how much energy goes into the sound waves, known as the efficiency, for which there is an input parameter in \texttt{BSMPT} which is the ratio of the efficiency of the sound wave contribution to the efficiency of the turbulent part. This ratio is taken to be \texttt{0.05} while the other efficiency factors are calculated by \texttt{BSMPT} \cite{Basler:BSMPT3}.

We also consider the expected signal-to-noise (SNR) at possible experiments. It provides a measure of the expected GW signal in comparison to the GW background. This quantity is defined by 
\begin{equation}
    \text{SNR}= \sqrt{\mathcal{T} \int_{f_{min}}^{f_{max}}df\left(\frac{h^2\Omega_{\rm GW}}{h^2\Omega_{\rm Sens}}\right)^2}.
\end{equation}
Here $h^2\Omega_{GW}$ is the predicted GW spectrum, $f_{\min}$ and $f_{\max}$ define the frequency range of the experiment, $\mathcal{T}$ is the observing time, and $h^2\Omega_{\rm Sens}$ is the relevant effective sensitivity curve including the assumed background treatment. For LISA we use the sensitivity prescription of Ref.~\cite{Caprini:LISADetetability}. Although the precise detection threshold depends on the search strategy and foreground assumptions, an SNR above $\sim10$ is commonly used as an indicative observability criterion \cite{Caprini:LISADetetability,Babak:LISASNR}.

\section{Parameter Scan}\label{sec::Scan}
To study the N2HDM we perform a scan over its parameter space, in both the conformal and non-conformal cases. In short, we use the masses and mixing angles as parameters and we use the same ranges in both models when applicable. For details of the scan we refer to appendix~\ref{sec::AppendixSearch}. Constraints are imposed to ensure that the points found are not unphysical or ruled out by experiment. The theoretical constraints are introduced in Sec.~\ref{sec::TheoConstraints} while the phenomenological ones are in Sec.~\ref{sec::PhenoConstraints}. 

\subsection{Theoretical Constraints} \label{sec::TheoConstraints}
To restrict the scan to physically meaningful and perturbatively controlled regions, we impose a set of theoretical constraints. In this section we briefly discuss the constraints used and show their impact on the parameter space.

\textbf{Electroweak Symmetry Breaking}: A necessary criterion for EWSB is that the minimum of the effective potential at the VEV is below the extremum at the origin. This condition can be written as: 
\begin{equation}
    V^{(1D)}(\phi=0)>V^{(1D)}(\phi=v_h),\label{eq::EWSBCondition}
\end{equation}
where $V^{(1D)}(\phi)$ is a one dimensional slice of the effective potential along the direction of the VEV. Using a similar analysis as performed by Gildener and Weinberg~\cite{Gildener:Conformal}, leads to the following condition for the SI model: 
\begin{equation}
    \sum_{i\in S} m_{i}^4+3\sum_{i\in G} m_{i}^4-4\sum_{i\in F}n_{c_i} m_{i}^4< 16\pi ^2m_{h_1}^2v_h^2.\label{eq::massLimit}
\end{equation}
Here $n_{c_i}$ is a factor for the number of degrees of freedom from color, $m_{h_1}$ is the
scalon mass at one-loop and the sums over $S$, $F$ and $G$ refer to summing over all scalars, fermions, and gauge bosons respectively. This leads to a bound on the masses of the BSM scalar particles, depending on the mass of $h_1$. However, as $v_h$ contains contributions from the singlet field, this restriction is less constraining than the case of the 2HDM.

Conditions~\ref{eq::EWSBCondition}--\ref{eq::massLimit} express the requirement that the electroweak minimum lie below the origin, evaluated on the one-loop effective potential in the modified on-shell scheme of Appendix~\ref{sec::AppendixCT}. Because that scheme fixes the scalon mass $m_{h_1}$ on-shell while the remaining scalar masses are scanned independently, \ref{eq::massLimit} does not coincide with the Gildener--Weinberg scalon-mass relation $\sum_i n_i(-1)^{2s_i}m_i^4 = 8\pi^2 v_h^2 m_{h_1}^2$: under that relation \ref{eq::massLimit} would reduce to $m_{h_1}^2>0$ and hold automatically, so it is the independent on-shell value of $m_{h_1}$, together with the freely scanned masses, that makes it a genuine constraint. We further note that Eq.~\ref{eq::EWSBCondition} compares the electroweak extremum with the origin along a single field direction and does not by itself establish that this extremum is the global minimum of the full $(\phi_1,\phi_2,\phi_S)$ potential.

For the full model, with no SI, there are also contributions from the quadratic terms in the potential, but the contribution from one-loop $m_{h_1}$ contributions disappears, which leads to 
\begin{equation}
\sum_{i\in S} m_{i}^4+3\sum_{i\in G} m_{i}^4-4\sum_{i\in F}n_{c_i} m_{i}^4 <   {4 \pi^2} \left (-m_{11}^2v_1^2-m_{22}^2v_2^2-m_S^2v_s^2+2m_{12}^2v_1v_2 \right ).
\end{equation}

\textbf{Boundedness from below}: The potential is required to be bounded from below (BFB) to ensure stability of the field centered around the VEV. Therefore, as any field goes to infinity, we enforce $V(\vec\phi)\geq 0$ leading to conditions on the quartic couplings. In brief, the general idea~\cite{Arhrib:N2HDMPheno} is to rewrite the potential in terms of
\begin{equation}
    r^2=H^\dagger_1H_1+H_2^\dagger H_2+S^2,
\end{equation}
as well as three angles $\gamma$, $\psi$, $\theta$ and to ensure that, as $r\to \infty$, all coefficients are positive for the allowed values of $\gamma$, $\psi$, $\theta$. This results in the following conditions: 
\begin{equation}
\begin{split}
    \lambda_1,\lambda_2,\lambda_6&>0,\\
    \lambda_3+\sqrt{\lambda_1\lambda_2}&>0,\\
    \lambda_3+\lambda_4-|\lambda_5|+\sqrt{\lambda_1\lambda_2}&>0,\\
    \lambda_7+\sqrt{\lambda_1\lambda_6}&>0,\\
    \lambda_8+\sqrt{\lambda_2\lambda_6}&>0.
\end{split}    
\end{equation}
If $\lambda_7$ or $\lambda_8$ are negative there are two additional criteria which need to be satisfied:
\begin{equation}
    \begin{split}
        \lambda_3\lambda_6-\lambda_7\lambda_8+\sqrt{(\lambda_1\lambda_6-\lambda_7^2)(\lambda_2\lambda_6-\lambda_8^2)}>0,\\
        \lambda_6(\lambda_3+\lambda_4+|\lambda_5|)-\lambda_7\lambda_8+\sqrt{(\lambda_1\lambda_6-\lambda_7^2)(\lambda_2\lambda_6-\lambda_8^2)}>0.
    \end{split}
\end{equation}
These inequalities define the region of parameter space where the potential remains BFB in all field directions and thus satisfy the constraint. Satisfying our BFB constraint corresponds to a parameter point having a local electroweak minimum; a dedicated check that this minimum is global — against competing singlet, inert-like or charge-breaking extrema — is beyond the scope of the present work and could exclude a subset of points.

\textbf{Perturbative unitarity}: Two necessary additional conditions are perturbativity and tree-level unitarity of the S-matrix, together called perturbative unitarity. Perturbativity is related to the size of the coupling parameters and is necessary for the treatment of the effective potential to be valid. This condition is enforced as 
\begin{equation}
    |\lambda_i|<4\pi.
\end{equation}

The unitarity of the S-matrix comes from studying the tree-level $2\to 2$ processes for scalars in terms of the scattering matrix. The eigenvalues of the matrix are required to have a magnitude less than $8\pi$ for the unitarity condition. The resulting conditions on the quartic couplings are, in the N2HDM, worked out in Ref.~\cite{Arhrib:N2HDMPheno} as 
\begin{equation}
    \begin{split}
        \left|\lambda_3+\lambda_4\right |<8\pi, \quad
        \left|\lambda_3\pm \lambda_5\right|<8\pi, \quad
        |\lambda_3+2\lambda_4\pm 3\lambda_5|&<8\pi,\\
        \left|\frac{\lambda_6}{4}\right|<8\pi,\quad \left|\frac{\lambda_7}{2}\right|<8\pi,\quad \left|\frac{\lambda_8}{2}\right|&<8\pi,\\
        \left|\frac{1}{2}\left(\lambda_1+\lambda_2\pm \sqrt{(\lambda_1-\lambda_2)^2+4\lambda_4^2}\right)\right|&<8\pi,\\
        \left|\frac{1}{2}\left(\lambda_1+\lambda_2\pm \sqrt{(\lambda_1-\lambda_2)^2+4\lambda_5^2}\right)\right|&<8\pi.
    \end{split}
\end{equation}
There are three additional eigenvalues that do not have a simple solution and instead the magnitude of the roots of the polynomial
\begin{equation}
    \begin{split}
    &2 x^3 + (-6\lambda_1- 6\lambda_2- \lambda_6) x^2\\ & + (18\lambda_1\lambda_2- 8\lambda_3^2- 8\lambda_3\lambda_4- 2\lambda_4^2 + 3\lambda_1\lambda_6 + 3\lambda_2\lambda_6- \lambda_7^2- \lambda_8^2 ) x\\
    &- 9\lambda_1\lambda_2\lambda_6 + 4\lambda_3^2\lambda_6 + 4\lambda_3\lambda_4\lambda_6 + \lambda_4^2\lambda_6 + 3\lambda_2\lambda_7^2- 4\lambda_3\lambda_7\lambda_8- 2\lambda_4\lambda_7\lambda_8 + 3\lambda_1\lambda_8^2 = 0,
\end{split}
\end{equation}
are required to be smaller than $8\pi$.

\begin{figure}[t]
    \centering
    \includegraphics[width=\linewidth]{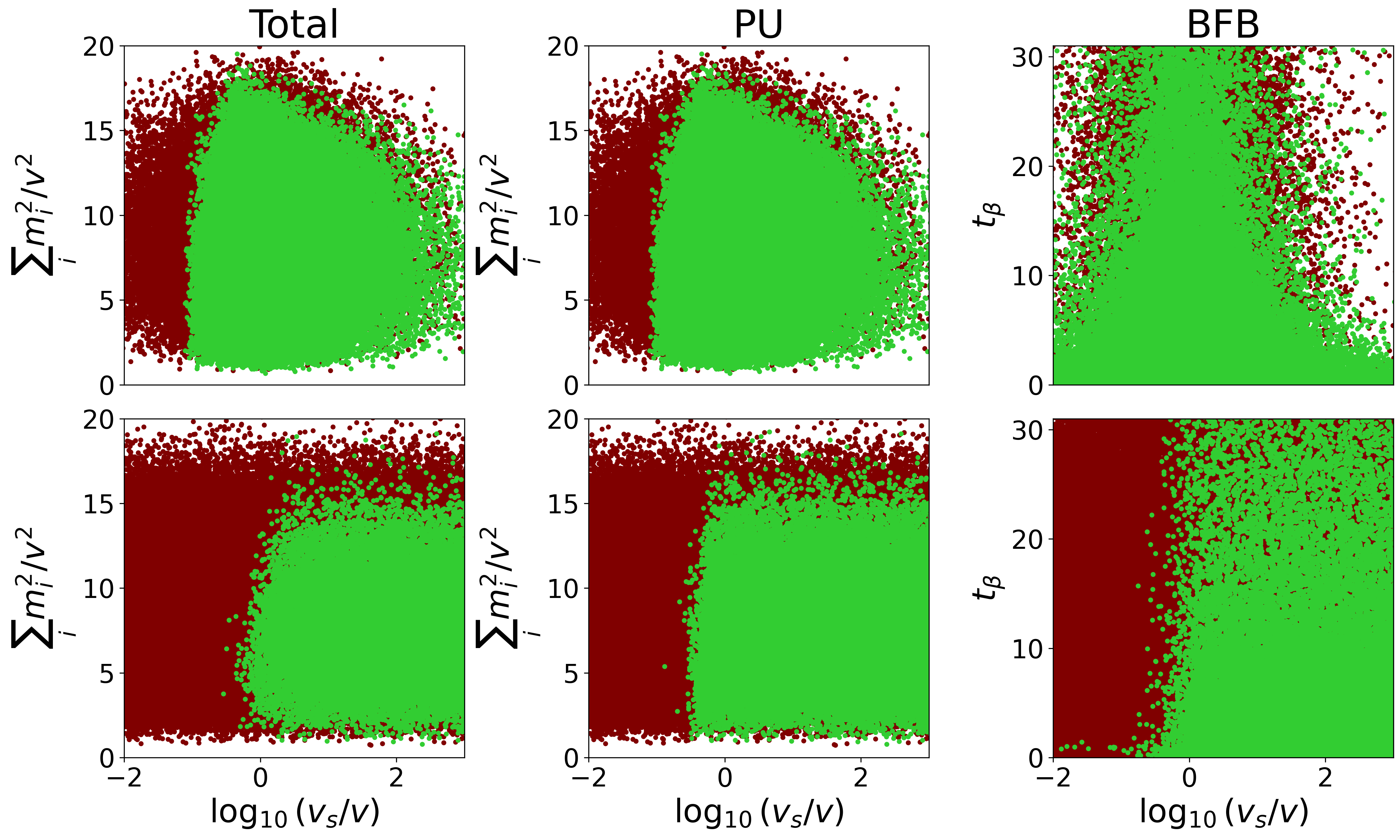}
    \caption{Upper (lower) panels show the theoretical constraints for the conformal (non-conformal) parameter space. The different columns show different constraints; from left to right: all theoretical constraints, perturbative unitarity and boundedness from below. The sum of masses squared refers only to the physical scalars. Points passing the constraints (in green) are plotted on top of points failing (red).}
    \label{fig:Theoretical Constraints}
\end{figure}

The impact of the theoretical constraints on the sum of the masses squared for all the physical scalars as well as on $t_\beta$ as function of the singlet VEV is shown in Fig. \ref{fig:Theoretical Constraints}, both for the conformal and non-conformal models\footnote{Note that the points plotted are a projection of a higher-dimensional parameter space onto a 2D surface. This means that even though the impact of some constraints are unclear in a certain projection it does not necessarily mean that they are irrelevant.}. 
From the figure it is clear that small singlet VEVs are not allowed when imposing all the theoretical constraints. This can be understood from the tadpole equations, Eq.~\ref{eq::tadpole1}, from which it follows that $\lambda_7$ has contributions that goes as $v^2/v_s^2$. In addition, it is also clear that the sum of scalar masses squared is limited by the perturbative unitarity, especially for the non-conformal model. The constraint from boundedness from below has a small effect on the conformal model whereas for the non-conformal model it gives a lower limit on the singlet VEV.

\subsection{Phenomenological constraints}\label{sec::PhenoConstraints}

In addition to theoretical constraints we also need to take into account experimental limits on the  Higgs particles predicted by the models. This is done by comparing the model's prediction for production rates with those observed at colliders. For each particle the most sensitive channel is chosen and a bound is put on the production rate using this channel. Specifically, for each particle, the chosen channel is enforced to have a $\chi^2$, relative to the SM, inside a 95 \% confidence interval, meaning $|\chi^2_{N2HDM}-\chi_{SM}^2|<6.18$. This method is used, rather than just dealing with $\chi^2_{N2HDM}$, as it leads to more lenient bounds in channels where the SM predicts the experimental signal poorly.

These constraints are implemented in the packages \texttt{HiggsBounds} \cite{Bechtle:HiggsBound1,Bechtle:HiggsBounds2,Bechtle:HiggsBounds4,Bechtle:Higgsbounds5} and \texttt{HiggsSignals} \cite{Bechtle:HiggsSignals1,Bechtle:HiggsSignals2}.
\texttt{HiggsSignals} provides experimental constraints on the signal of the SM-like Higgs that has been discovered. Meanwhile, \texttt{HiggsBounds} imposes limits on the BSM scalars to ensure that their predicted experimental signal does not violate known bounds. The programs use experimental data from LEP, Tevatron and the LHC to apply the bounds,  using $\chi^2$, in the manner explained earlier. 

In practice the constraints are applied using \texttt{ScannerS} \cite{Muhlleitner:N2HDM,Muhlleitner:ScannerS,Coimbra:ScannerSOrgiginal}, which serves as an interface to the \texttt{HiggsBounds}  and \texttt{HiggsSignals} packages. Internally, the \texttt{ScannerS} program also calculates the oblique parameters \cite{Peskin:Oblique} and uses the package \texttt{AnyDecay} \cite{Djouadi:HDecay1,Djouadi:HDecay2,Engeln:HDecay3} to   calculate the decay widths needed for \texttt{HiggsBounds} and \texttt{HiggsSignals}.

There are also constraints on the charged Higgs particle, $H^\pm$, from B-physics, such as $b\to s\gamma$, that place important bounds on the possible parameter space. This is implemented in \texttt{ScannerS} using a fit \cite{Haller:ObliqueData} of several experimental channels resulting in a $2\sigma$ exclusion bounds on the parameter space of $m_{H^\pm}$ and $t_\beta$.

\begin{figure}[!htb]
    \centering
    \includegraphics[width=0.9\linewidth]{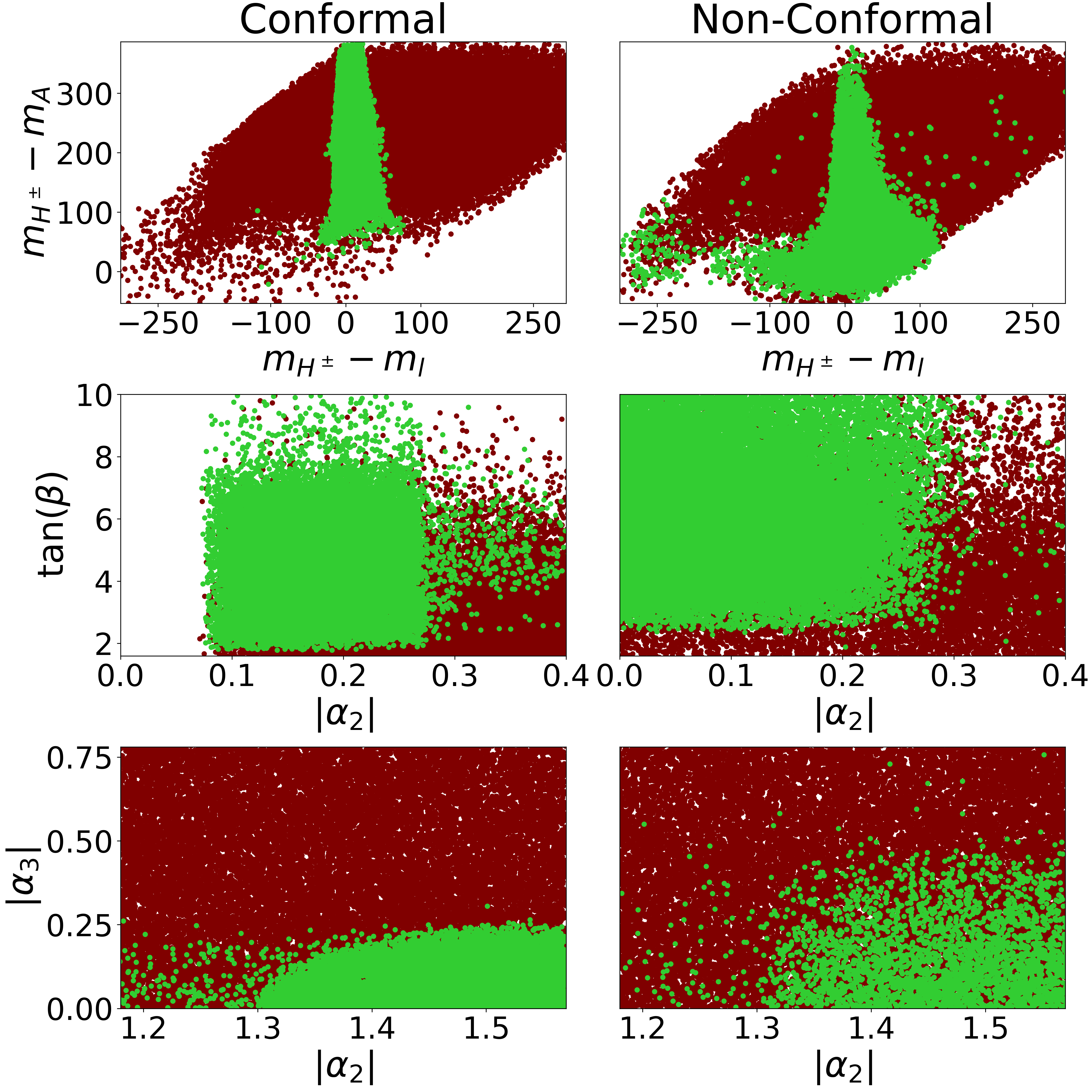}
    \caption{The effects of the phenomenological constraints on the conformal (left) and non-conformal (right) models. All points have passed the theoretical constraints where green (red) points passed (failed) the phenomenological constraints. The upper row shows the difference of masses (measured in GeV) between the charged scalar Higgs, the CP-odd Higgs and the lightest CP-even BSM Higgs, $l$. The middle row shows the rotation angle $\alpha_2$ and $\tan(\beta)$ in scenario A and the lower row shows the rotation angles $\alpha_2$ and $\alpha_3$ in scenario B.}
    \label{fig:PhenoConstraints}
\end{figure}

Figure~\ref{fig:PhenoConstraints} shows the impact of the phenomenological constraints on points that pass all theoretical constraints. 
The upper row shows the impact of the oblique parameters, enforcing the mass of $H^\pm$ to be near that of the CP-odd or lightest CP-even Higgses. For the conformal scenario, this results in the mass of the charged Higgs being near that of the lightest CP-even Higgs, $m_{H^\pm} \approx m_l$, when including all constraints. For the non-conformal model there are also points in the region $m_{H^\pm}\approx m_A$. The difference  arises from the collider phenomenology constraints, which do not allow points with $m_{H^\pm}\approx m_A$ in the conformal case.   

In addition to the constraints on the masses, there are also limitations on the mixing angles. The middle row in the figure shows $\alpha_2$ and $t_\beta$ for scenario A, as defined in Eq.~\ref{eq::scenarioA}. This shows that the magnitude of $\alpha_2$ is limited to be near alignment which corresponds to the $hVV$ couplings being close to the SM value. 
In addition there is also a lower limit on $\alpha_2$ in the conformal case that is not present in the non-conformal one. This can be understood 
from the theoretical constraints limiting $v_s\to0$, which impacts $\alpha_2$ in the conformal model through Eq.~\ref{eq::alpha2}.
The figure also shows that points with small values of $t_\beta$ struggle with experimental constraints due to large Yukawa couplings. The third row shows a similar phenomenon for scenario B where $\alpha_2$ is constrained to be close to alignment. In addition, $\alpha_3$ is similarly constrained to be small, which follows from the $hVV$ coupling for this scenario, see Tab.~\ref{tab:eff_couplings}. The tight constraints on the $\alpha_2$ mixing angle in both scenarios can be understood by the strong experimental bounds on $h\to VV$ decays \cite{ATLAS:2HDMConstraints}. 

\section{Electroweak Phase Transition Observables}\label{sec::EWPT}
The program \texttt{BSMPT} \cite{Basler:BSMPT1,Basler:BSMPT2,Basler:BSMPT3} is used to calculate the thermal parameters introduced in Sec.~\ref{sec::Thermal}. This program models the electroweak transition and calculates key properties such as the bounce, the latent heat released and the characteristic temperatures of the EWPT. To use the conformal model, the modified CTs following from 
Eq.~\ref{eq::ModifiedOS} were implemented in \texttt{BSMPT}, see appendix \ref{sec::AppendixCT} for formulas, whereas for the standard OS scheme we used the implementation in  \texttt{BSMPT}. 
With the necessary CTs implemented, the effective potential is defined and \texttt{BSMPT} was used to trace minima of the potential over a range of temperatures to determine whether a PT happens. For first-order transitions with successful tunneling and percolation, the program returns the transition parameters used for the baseline GW estimate.

In this section, we discuss our findings from applying \texttt{BSMPT} to the parameter points in the scan that fulfill the theoretical constraints 
for both the conformal and non-conformal model. In particular we investigate the impact of the phenomenological constraints  on the detectability of the GWs as well as the properties of the first-order PTs and how different masses and couplings of the models correlate with the GW observables. In addition to our general findings, we also showcase specific values of masses and thermal properties for four benchmark points in Tab. \ref{tab:BenchmarkPoints}. 

\begin{figure}[!htb]
    \centering
    \includegraphics[width=0.9\linewidth]{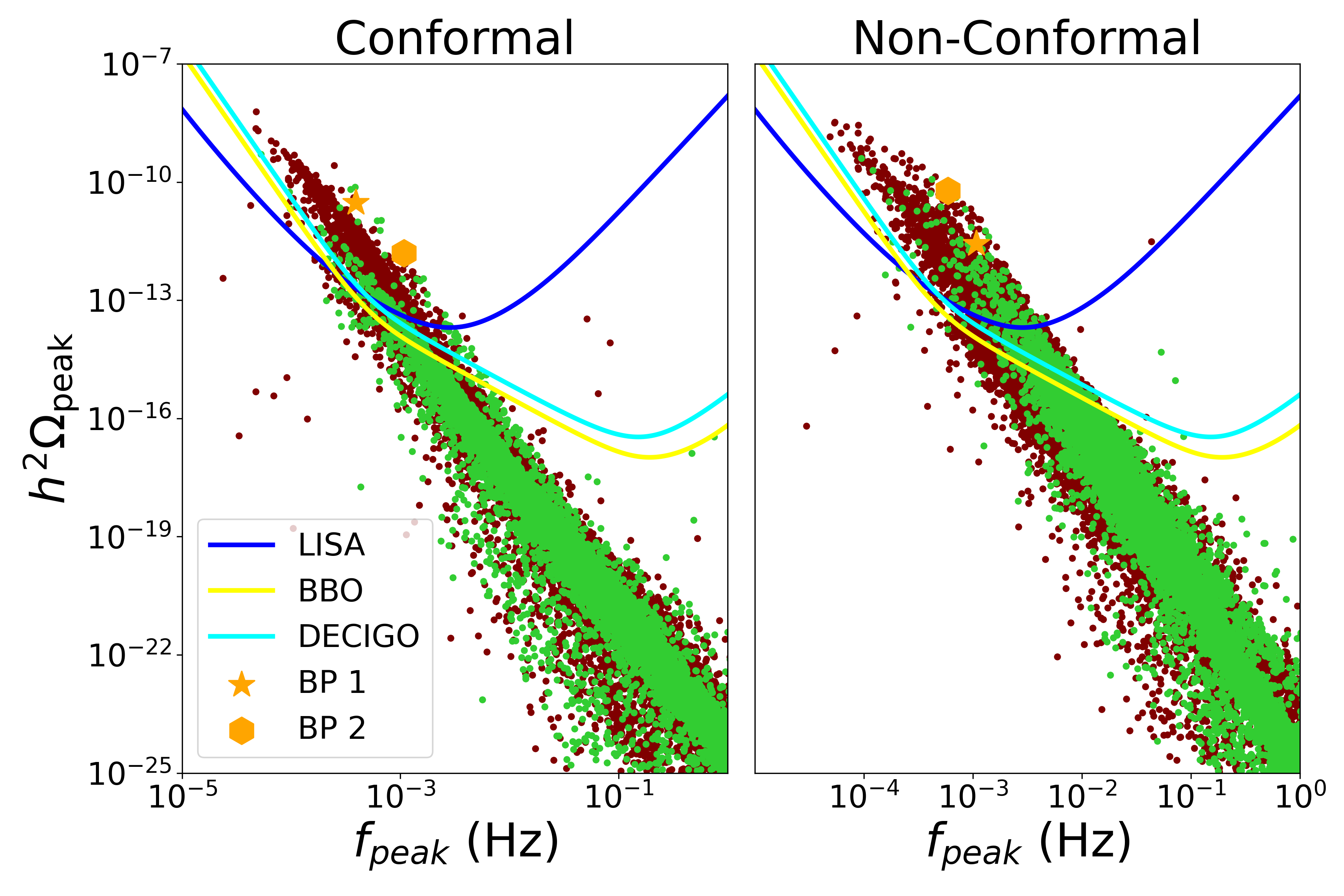}
    \caption{The power spectrum peak and the corresponding frequency for the conformal and non-conformal model. The color represents whether a model passed the phenomenological constraints imposed through \texttt{ScannerS} (green passes, red fails). All points passed the theoretical constraints.}
    \label{fig:PowerSpectrumPheno}
\end{figure}

As already emphasized, the results shown in this section correspond to the baseline sound-wave contribution, as calculated by \texttt{BSMPT}.
We do not claim these SNR values as precision forecasts. 
In particular, for very strong transitions, $\alpha \gtrsim 1$, the acoustic source may last for less than one Hubble time, which can suppress the sound-wave contribution \cite{Ellis:2018mja}. 
Possible contributions from turbulence and bubble-collisions, as well as the detailed wall dynamics, are not included point by point. 
The largest-SNR points should therefore be interpreted as optimistic indicators of detectability, while the relative comparison between conformal and non-conformal parameter regions is the more robust conclusion. 

The impact of phenomenological constraints on the power spectrum  of the GW, characterized by the peak frequency and strength, is shown in Fig. \ref{fig:PowerSpectrumPheno}. As can be seen from the figure, the constraints have an important impact on the region corresponding to energetic GWs. Specifically, when the power spectrum peak has a high value it is found that there are very few points that pass the phenomenological constraints in both the conformal and non-conformal models. It is also clear that the conformal and non-conformal models are similar to one another as both span a similar range of values for the power spectrum peak with only a small tail of strongly supercooled ($\alpha\gtrsim1$) transitions reaching the largest peak values. 

In addition to the results of the scan, the figures also show  sensitivity curves, from Ref.~\cite{Caprini:LISADetetability}, for the future LISA, BBO and DECIGO experiments. This roughly indicates in what region of parameter space experiments will be able to detect GW (with the regions above the curve indicated as detectable). This shows that, within the adopted sound-wave treatment and scan ranges, there are points in both scenarios that pass all imposed constraints and lie in a region that can be probed by future experiments. Of particular interest is the LISA experiment, as it is planned to be launched within the next decade, and we find points that could reasonably be detected by LISA in the N2HDM for both the conformal and non-conformal scenarios. 

\begin{figure}[!htb]
\centering
\includegraphics[width= \linewidth]{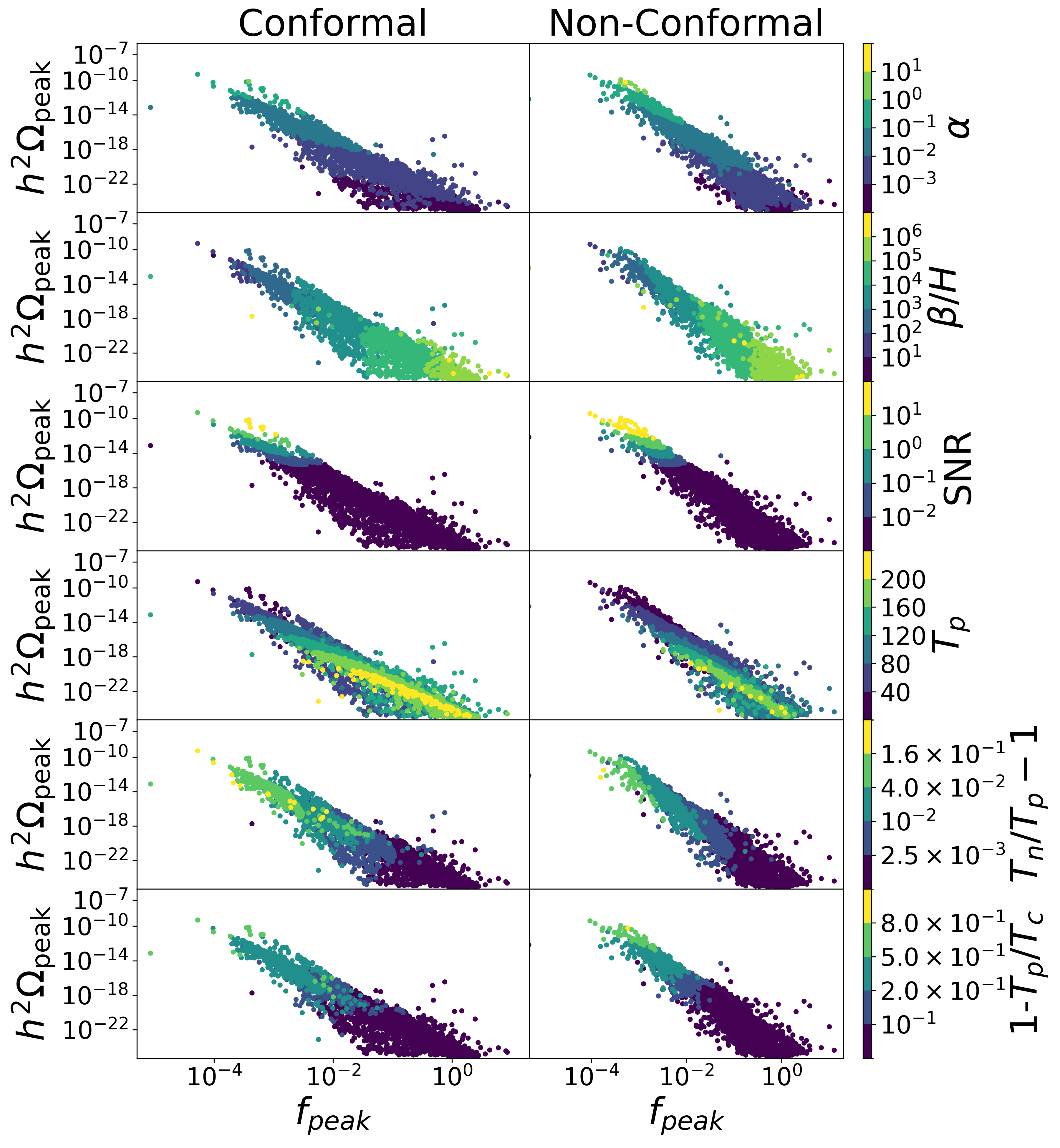}
\caption{The axes show the power spectrum peak and the corresponding peak frequency, while the colour indicates observables related to the electroweak phase transition. From top to bottom the observables are: the strength, $\alpha$, the characteristic time scale of the PT, $\frac{\beta}{H}$, the signal-to-noise ratio expected at LISA (assuming a 3 year operation time), the percolation temperature, $T_p$ (in GeV) and the quantity $T_n/T_p-1$ measures the delay between nucleation and percolation. Each point passed all theoretical and phenomenological constraints and has a completed transition according to the percolation criterion used in the scan.}
\label{fig:PTObservables}
\end{figure}

Apart from the power spectrum itself we also show in Fig.~\ref{fig:PTObservables} how the EWPT observables correlate with the peak frequency and strength. In both models the strength, $\alpha$, and characteristic time-scale, $\frac{\beta}{H}$ correlate with the power spectrum peak as expected from theory. Meaning, higher power spectrum peaks have larger $\alpha$ and smaller $\frac{\beta}{H}$, interpreted as the EWPT lasting longer. In addition, the detectability of the GW is shown through the SNR of LISA, calculated by \texttt{BSMPT} assuming a 3 year operation time, which is plotted in the third row. This verifies that the sensitivity curve gives a good estimation of the region of detectability as the region with a SNR larger than 10 agrees well with the sensitivity curve of LISA. From this, we conclude that there are points passing the phenomenological constraints whose predicted sound-wave signal would be detectable at LISA in both the conformal and non-conformal model.

In the last three rows of the figure, quantities related to the temperature of the EWPT are shown. First of all, the figure shows that the percolation temperature has a large range of values from around 10~GeV up to temperatures over 250~GeV. We also find that higher power spectrum peaks tend to have lower percolation temperatures. This means the EWPT ended at a lower temperature and therefore at a later time, which fits with $\frac{\beta}{H}$. 

Finally, the figure shows how the percolation temperature differs from the nucleation temperature as well as the critical temperature. Since $T_n\simeq T_p$ across the accepted sample, we find no significant delay between nucleation and percolation, and the transitions complete promptly. This does not exclude supercooling in the strict sense: the critical temperature $T_c$ can lie substantially above $T_n$ and $T_p$, so that a point may cool well below $T_c$ before nucleating while still  completing rapidly thereafter. We quantify the amount of supercooling by $1-T_p/T_c$, which ranges over $0\ldots0.8$ across the accepted points. The bulk of the sample, with weak-to-moderate transition strength ($\alpha\lesssim1$), shows only mild supercooling, whereas a tail of the strongest transitions ($\alpha\gtrsim1$, up to $\alpha\simeq6.5$ for the benchmark NC-BP2) reaches $1-T_p/T_c\simeq0.8$, i.e.\ substantial supercooling with the Universe cooling to $T_p\simeq0.2\,T_c$ before the transition completes. These are the same points for which the percolation temperature is lowest and the power-spectrum peak highest, and for which the sound-wave-only SNR should be read with the caveats of Sec.~\ref{sec::Thermal}. What is robust across the whole sample is therefore the prompt completion of the transition ($T_n\simeq T_p$) rather than the absence of supercooling. A similar pattern---mild supercooling for most points with a strongly-supercooled tail---is found in the \texttt{ScannerS}-constrained N2HDM analysis of Ref.~\cite{Lee:2HDMNoSupercooling}; a direct comparison with Ref.~\cite{Benincasa:2HDMSupercooledConformal} is complicated by their more lenient bounds and different ($\overline{\textrm{MS}}$) renormalization scheme.

For reference, we also include the critical temperature which shows that while the nucleation and percolation temperature remain close, both can differ substantially from the critical temperature. This occurs especially in the case of strong phase transitions, $\alpha \gtrsim 1$, when the percolation temperature is low and the GW power spectrum peak is high.

\begin{figure}[!tb]
    \centering
    \includegraphics[width=\linewidth]{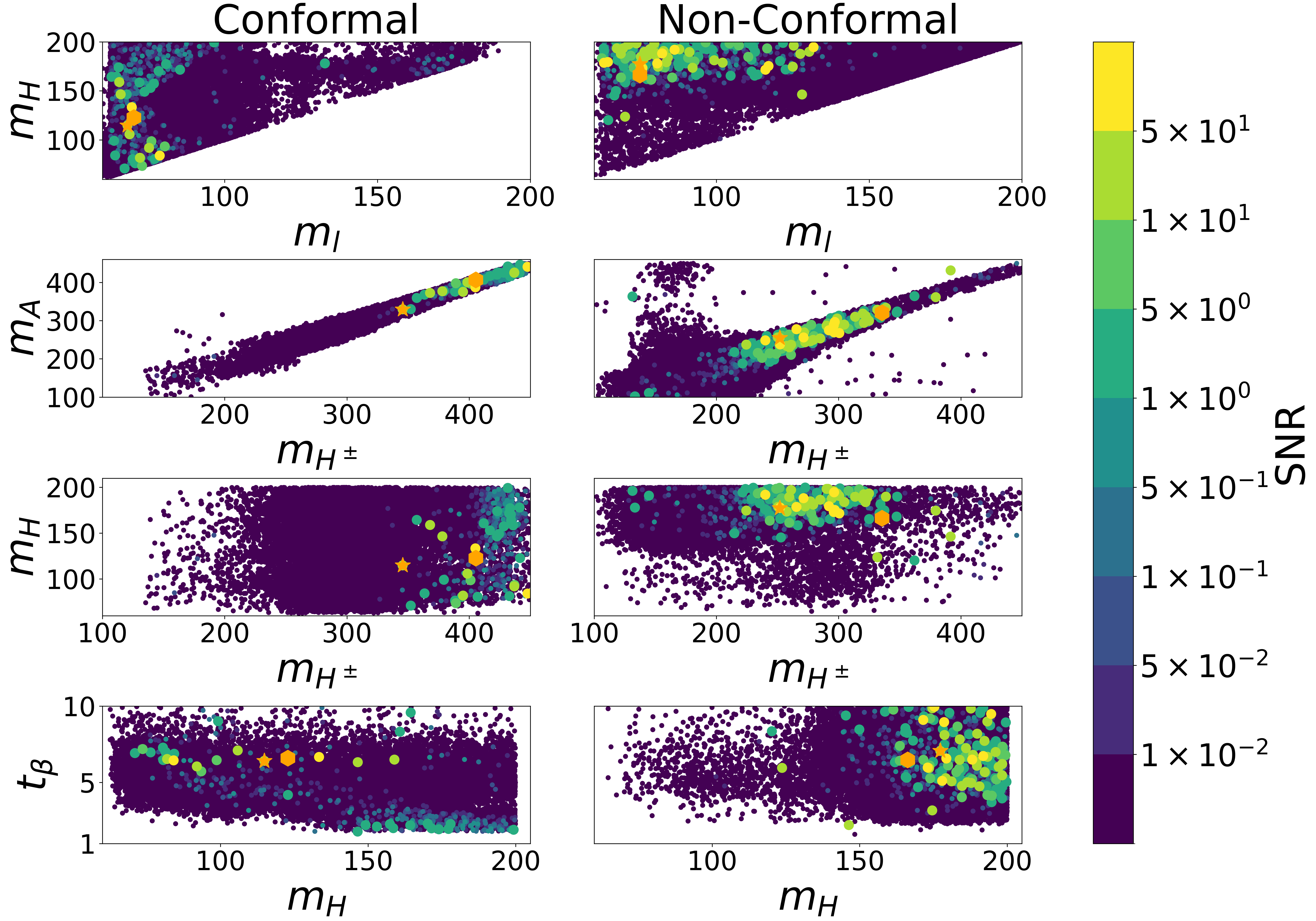}
    \caption{Color indicates the signal-to-noise ratio for LISA (assuming a 3 year operation time), for both the conformal and non-conformal scenario, where the largest values are plotted on top. The axes show model parameters, such as masses of charged and CP-odd Higgs, in order to show a difference between the conformal and non-conformal model. }
    \label{fig:ConformalDeliniation}
\end{figure}

Next we turn to the correlation of the LISA 3-year SNR with masses and couplings of interest for collider phenomenology as shown in 
Figure \ref{fig:ConformalDeliniation}. These observables 
were chosen as they, in combination with the observation of GWs, can be used to distinguish between the conformal and non-conformal models. In other words, if stochastic GWs and masses of  additional Higgs particles were to be measured these observations could give a strong indication of whether they correspond to a conformal model or not. 

As an illustration of how these observables delineate between types of model, the first three rows in the figure shows the masses of the charged and CP-odd particles as well as of the two additional CP-even Higgs particles. From these panels it is clear that the regions of large SNR, and therefore observable GWs, are in different mass ranges for the two models. As an example, in the conformal model the charged Higgs mass is large, above $\sim350$~GeV for observable GWs, while it is smaller in the non-conformal model, less than $\sim350$~GeV. We also see a similar correlation for the CP-odd Higgs. The third row shows the mass of $H$ and $H^\pm$, finding similarly that the region of observability for GWs are distinct when comparing across the type of model. Detectable GWs are, in the sampled set, found for large $m_H$, larger than $150$~GeV, in the non-conformal case and for a large range of values in the conformal case.

In addition to the masses, we also look at $t_\beta$, or the VEV structure, in the lowest row of the figure. This shows, similarly to the other panels, that the regions of GW observability is distinct between models in the $t_\beta$-$m_H$ plane. The non-conformal points of large SNR are found for $m_H>150$~GeV and $t_\beta>3$ whereas the detectable GWs in the conformal model are not found in this region. This gives another, scan-level handle for using collider measurements in combination with gravitational waves to discriminate between conformal and non-conformal N2HDM realizations. It should be noted that the number of high SNR points is $\sim 50$ so these conclusions should be seen within the context of our scan of the ranges of parameters we have used.

\begin{table}
    \centering
\caption{
Representative benchmark points from the conformal and non-conformal scans. The quoted transition and GW quantities are evaluated using the same temperature convention and wall velocity as described in Secs.~\ref{sec::Thermal} and~\ref{sec::EWPT} The SNR values correspond to the baseline LISA sound-wave estimate for a three-year observation time and should be interpreted with the source-lifetime caveats discussed in the text.
}   
\vspace*{0.2cm}
    \begin{tabular}{|c|c|c|c|c|}
         \hline 
         Observable & C-BP1 &C-BP2&NC-BP1&NC-BP2\\
         \hline 
         $m_{l}$ [GeV] & 68& 70&75&75 \\
         $m_{H}$ [GeV] & 115&123&177&166\\
         $m_{H^\pm}$ [GeV]& 346&405&252 &335\\
         $m_{A}$ [GeV] & 330&407&254&321\\
         $t_\beta$&6.4&6.6&7.1&6.5\\
         $v_s$ [GeV] & 59&805&202&245\\
         $T_n$ [GeV]& 31&36&27&12\\
         $T_p$ [GeV] &29&35&26&12\\
         $\alpha$&0.73&0.25&0.61&6.5 \\
         ${\beta}/{H}$&260&610&840&1020 \\
         SNR & 42&14&23&190\\
         \hline          
    \end{tabular}
    \label{tab:BenchmarkPoints}
\end{table}

Finally we note that there are some points, such as the benchmark point NC-BP2, in Tab. \ref{tab:BenchmarkPoints}, that has a comparatively large value of $\alpha$. For such points the sound-wave-only SNR should be regarded as indicative rather than as a precision forecast, since acoustic-source lifetime effects and possible non-sound-wave contributions can be quantitatively important. This caveat does not affect the qualitative conclusion that both conformal and non-conformal scans contain regions with potentially observable stochastic backgrounds.

\section{Higgs-pair Production Complementarity}\label{sec::DiHiggs}
\begin{figure}[!htb]
    \centering
    \includegraphics[width=0.8\linewidth]{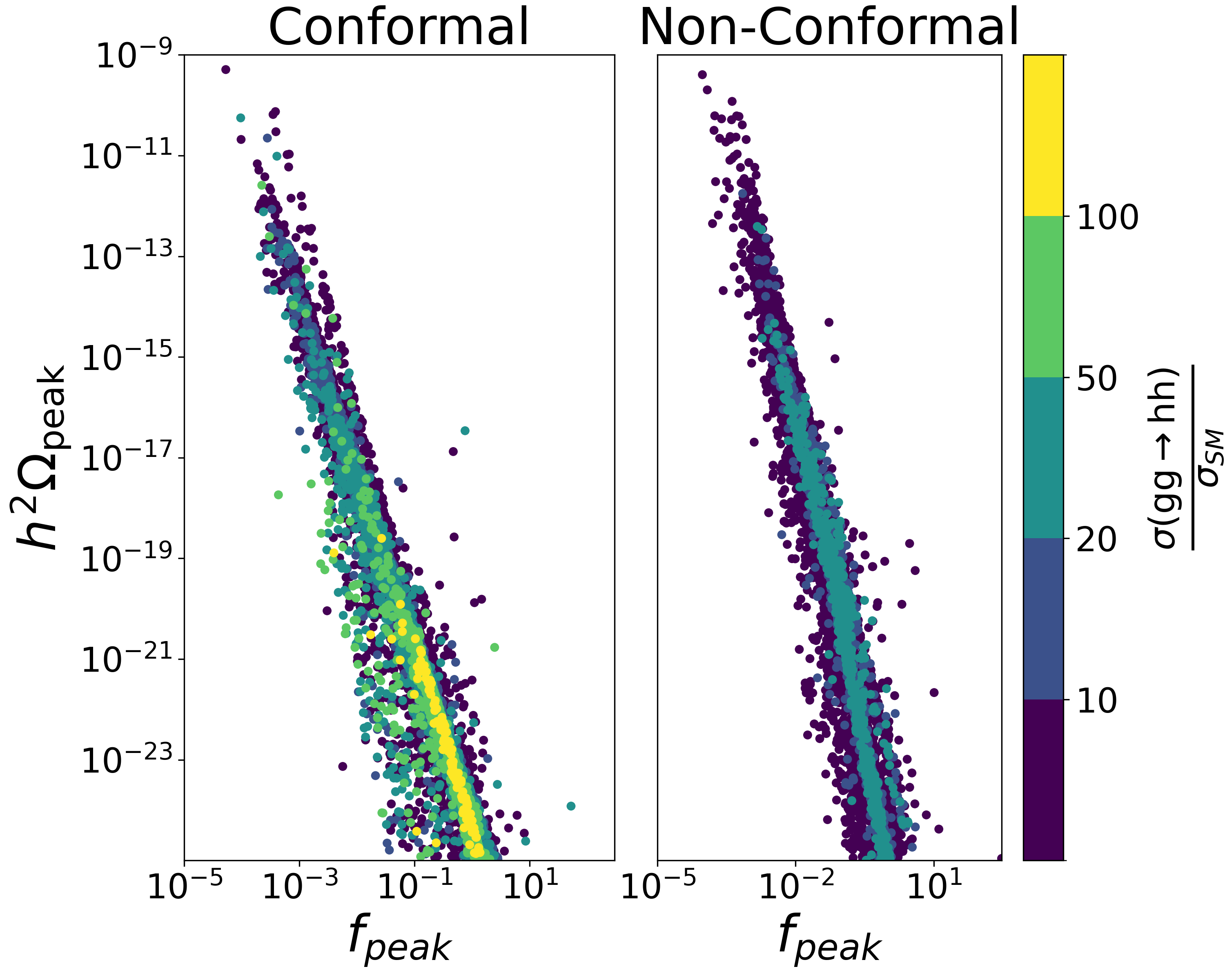}
    \caption{The left (right) panel shows the power spectrum in the conformal (non-conformal) model. The color indicates the fixed-energy partonic gluon-fusion rate at $\sqrt{\hat s}=400$~GeV into two SM-like Higgs bosons, normalized to the corresponding SM partonic rate in the same heavy-top approximation.}
    \label{fig:DiHiggsPowerSpectrum}
\end{figure}

\begin{figure}[!htb]
    \centering
    \includegraphics[width=\linewidth]{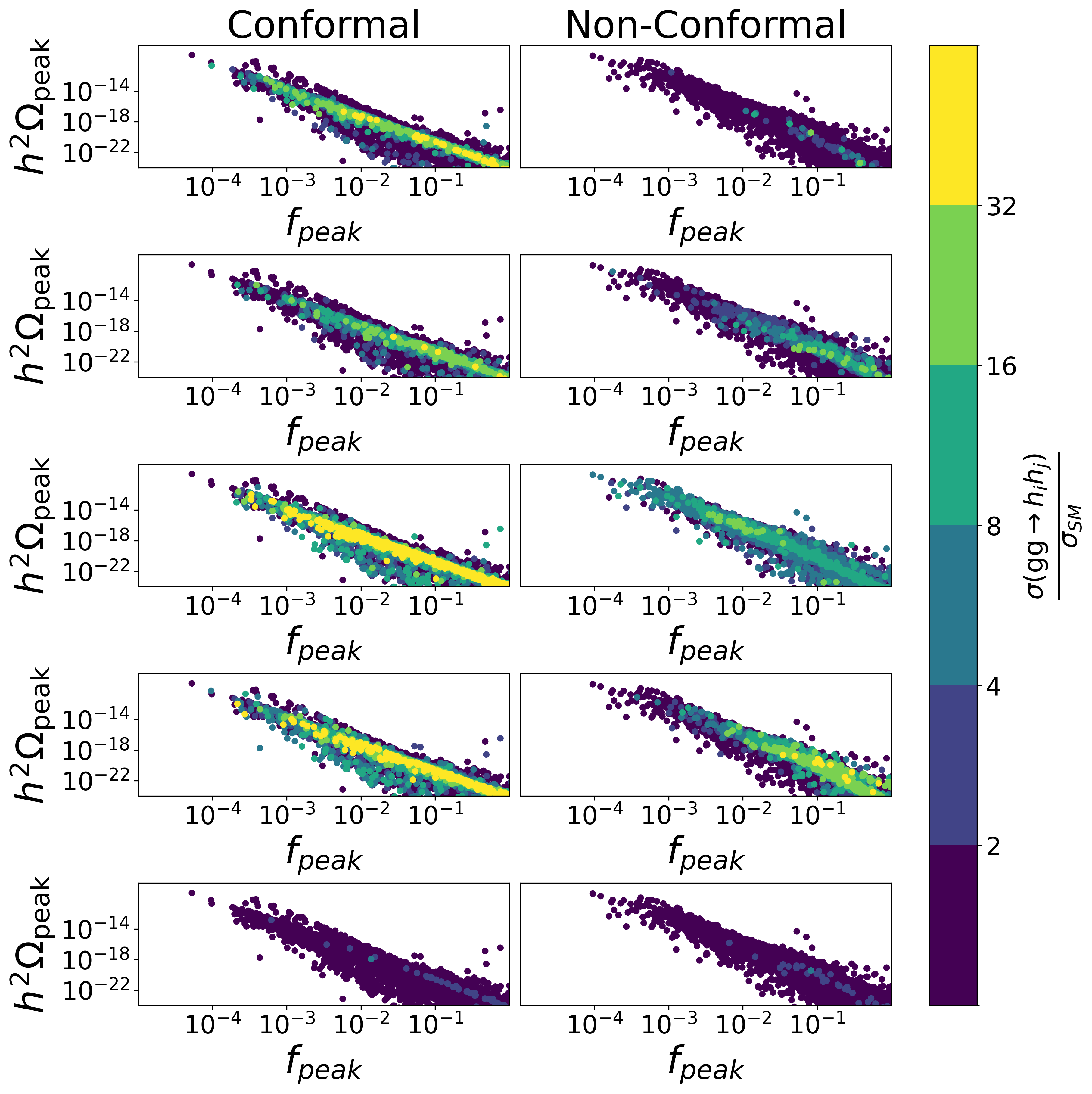}
    \caption{The power spectrum for the conformal and non-conformal scenario showing partonic cross sections for the di-Higgs production of BSM CP-even Higgs particles in color. The color is given by the partonic cross section of the production divided by the SM value of the di-Higgs process. From up to down the final state is: $ll$, $HH$, $hl$, $hH$ and $lH$.}
    \label{fig:BSMDiHiggsProduction}
\end{figure}

Higgs-pair production at hadron colliders, such as the upcoming HL-LHC, will present another possibility to probe the scalar potential of models with an extended Higgs sector. For simplicity, we study a fixed-energy partonic proxy\footnote{This is a partonic diagnostic rather than a hadron-collider prediction: it does not include parton distribution functions, finite top-quark-mass effects, decays or acceptance cuts. The chosen $\sqrt{\hat s}=400$~GeV lies just above the $t\bar t$ threshold $2m_t\approx346$~GeV, where the heavy-top approximation is least reliable, and near several scalar-pair thresholds; the zero-momentum effective trilinears employed here likewise omit momentum dependence and are not equivalent to on-shell vertices. We therefore refrain from drawing quantitative HL-LHC reach conclusions from Figs.~\ref{fig:DiHiggsPowerSpectrum} and \ref{fig:BSMDiHiggsProduction}.} for gluon-induced Higgs-pair production, $gg\to h_ih_j$, evaluated at a representative partonic centre-of-mass energy $\sqrt{\hat s}=400$~GeV in the heavy-top-quark limit. Thus,  
this section should only be interpreted as a partonic-level diagnostic of Higgs-potential complementarity, not as a full collider-reach analysis. A direct LHC prediction would require convolution with parton distribution functions, finite-top-mass effects, higher-order QCD corrections, scalar branching ratios and experimental selections. For the purpose of this paper, the normalized quantities shown below are nevertheless useful for identifying where in the parameter space enhanced scalar self-interactions occur and to what extent it correlates with the production of primordial GWs.

After EWSB, terms of the form
\begin{equation}
    \lambda_{ijk} h_i h_j h_k,
\end{equation}
will appear in the potential, which are important to di-Higgs production. The values of $\lambda_{ijk}$ will depend on the quartic couplings in the Lagrangian, which allows di-Higgs production to provide an indication of the shape of the Higgs potential around the minima. In turn, this provides additional information about the potential to what is already known from the mass of the Higgs. To study these terms we use the $\kappa$ parameters, defined as 
\begin{equation}
    \kappa_{ijk}=\frac{\lambda_{ijk}}{\lambda_{SM}},
\end{equation}
with $\kappa_\lambda=\kappa_{111}$ defined to be the SM-like Higgs trilinear coupling to itself normalized to the SM value $\lambda_{SM}$. This parameter has a current experimental bound of $-1.3<\kappa_\lambda<6.7$ \cite{CMS:2026nuu} (at 95\% CL) while in future this bound is expected to shrink to $0.5<\kappa_\lambda<1.7$ \cite{ATLAS:2025eii} at the high-luminosity LHC upgrade. These limits are obtained assuming that the Higgs-pair production is close to being SM-like. In the N2HDM, additional scalar propagators and modified Yukawa couplings can change the mapping between $\kappa_\lambda$ and the measured cross section and thus, these limits cannot be used directly but rather as an indication of allowed values.

The trilinear couplings are calculated at the one-loop level, assuming $p=0$ for all external momenta, by \texttt{BSMPT}. The program does this by taking derivatives of the effective potential, using expressions from Ref.~\cite{Camargo-Molina:PotentialDerivative}, which gives the couplings. These are used in our calculation in place of the tree-level trilinears as they give a more accurate estimate of the couplings, even though it uses zero external momentum for all particles involved. 

From the trilinear couplings and the top Yukawa couplings we can then calculate the partonic cross section for gluon-gluon fusion in the heavy-top approximation using the diagrams:

\begin{center}
\includegraphics{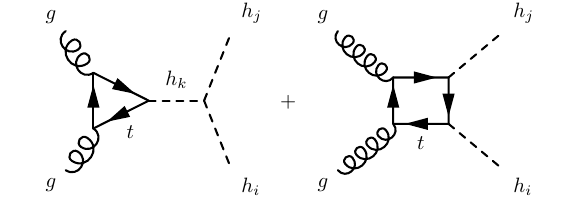}
\end{center}

In the heavy-top limit, the form factors that enter the relevant Feynman diagrams are given in Refs.~\cite{Plehn:DiHiggs,Dawson:DiHiggs} which allows for evaluation of the partonic cross as:
\begin{equation}
\hat{\sigma}_{\mathrm{LO}}(gg \to h_i h_j)=\int_{\hat t_-}^{\hat t_+} d\hat t\;\frac{G_F^{\,2}\alpha_s^{2}(\mu)}{256(2\pi)^3}\left|C_{ij}(\hat s)+D_{ij}\right|^2,\label{eq::dihiggs_sigma_corrected}
\end{equation}
where the triangle and box contributions are given by
\begin{equation}
    C_{ij}(\hat s)=\sum_{k=1}^3 \kappa_{ijk}\,c_{h_ktt} \,\frac{3m_h^2}{\hat s-m_{h_k}^2+i m_{h_k}\Gamma_{h_k}},\qquad
    D_{ij}=-\frac{4}{9}\,c_{h_itt}c_{h_jtt}.\label{eq::dihiggs_formfactors_corrected}
\end{equation}
Here $c_{h_itt}$ denotes the top-Yukawa coupling of $h_i$ normalized to the SM Higgs coupling, $\kappa_{ijk}$ is the normalized one-loop trilinear coupling calculated by \texttt{BSMPT}. Finally, the integration limits are
\begin{equation}
    \hat t_{\pm}=\frac{1 }{2}\left(m_{h_i}^2+m_{h_j}^2-\hat s \pm\sqrt{(m_{h_i}^2+m_{h_j}^2-\hat s)^2-4m_{h_i}^2m_{h_j}^2}
   \right)
.\label{eq::dihiggs_tlimits_corrected}
\end{equation}
If $i=j$ in Eq. \ref{eq::dihiggs_sigma_corrected} the final states are identical and the phase space integral needs to be taken more carefully. To ensure we only integrate over the parts of the phase space corresponding to a unique final state we then set the lower bound of the integration limit to correspond to $c_\theta=0$ and the upper one to $c_\theta=1$, where $\theta$ is the polar angle. If the final states are distinct we integrate over the whole phase space from $c_\theta=-1$ to $c_\theta=1$, using the $\hat{t}_\pm$ as our bound. 

In the following we will only consider the normalized partonic cross sections, 
\begin{equation}
\hat{\sigma}_{\mathrm{LO}}(gg \to h_i h_j) / \hat{\sigma}_{\mathrm{LO}}^{\mathrm{SM}}(gg \to h h)
\end{equation}
which should be viewed as a diagnostic to compare scan points rather than as a precision hadronic prediction.
Figure \ref{fig:DiHiggsPowerSpectrum} shows how the GW power spectrum and the di-Higgs production of SM-like Higgses complement each other. It shows that there are points with very large values for the partonic cross section diagnostic, however, these are mostly found for small values of the power spectrum peak. For larger values of the power spectrum peak, which could be detected by LISA, the partonic cross section is typically closer to the SM value, especially in the non-conformal case. Even so, it is clear that in the conformal model there are some points in this region with the partonic cross section diagnostic being larger than 10. Therefore, if the HL-LHC were to find a significant deviation in di-Higgs production and at the same time LISA would find energetic primordial GWs this would point to a conformal model. At the same time, even if no significant deviation was found at the HL-LHC, this would not rule out the possibility to find energetic primordial GWs.

\begin{figure}[!htb]
    \centering
    \includegraphics[width=0.8\linewidth]{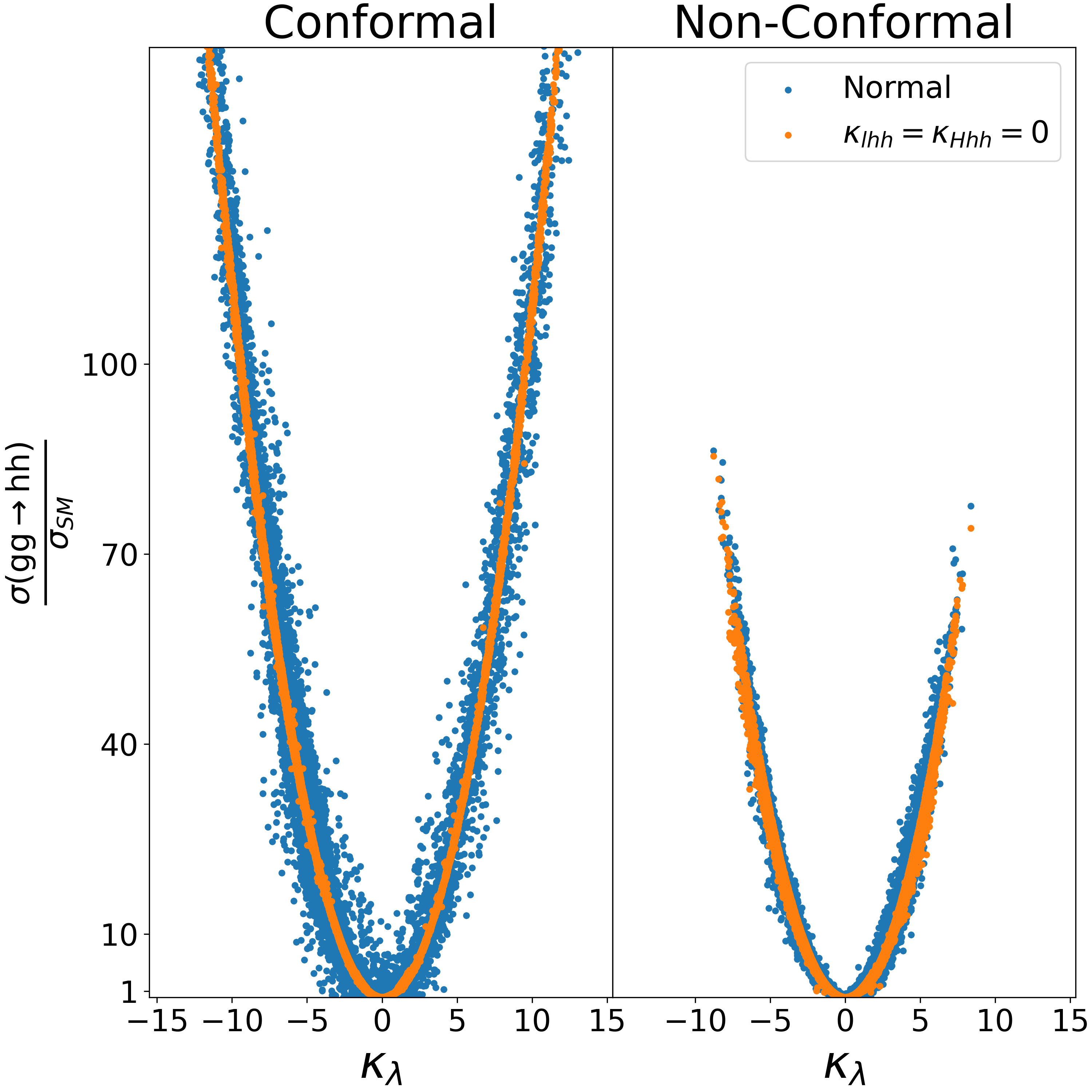}
    \caption{Fixed-energy partonic di-Higgs rate as a function of the trilinear coupling $\kappa_\lambda$ in the heavy-top approximation for the conformal and non-conformal scenarios. These rates are for the production of two SM-like Higgs bosons and are normalized to the corresponding SM partonic value at $\sqrt{\hat s}=400$~GeV. The orange points were obtained by setting all non-SM trilinear contributions to zero, leaving only the SM-like self-coupling contribution. Both panels use the same axis ranges so that the difference in the accessible range of $\kappa_\lambda$ between the two models can be read off directly.}
    \label{fig:DiHiggsTrilinear}
\end{figure}

A large partonic cross-section for di-Higgs production is to a large extent driven by the trilinear coupling of the SM-like Higgs, $\kappa_\lambda$, as is shown in Figure \ref{fig:DiHiggsTrilinear} for both the conformal and non-conformal models. To illustrate this further, the figure also shows the results when setting non-SM couplings to zero. As is clear from the figure, in the non-conformal case the effect of the BSM Higgs particles is very small whereas it can be quite significant in the conformal case, especially for small values of $\kappa_\lambda$. This can be understood from the quartic couplings typically being larger in the conformal case, in order to get Higgs masses in the range considered, since there is no contribution to the physical masses from mass terms in the Higgs potential. Within the scanned ranges the non-conformal points reach only $|\kappa_\lambda|\lesssim10$, whereas the conformal ones extend to $|\kappa_\lambda|\approx15$, consistent with the larger quartic couplings
in the conformal case.

To further illustrate the possibility of probing other quartic couplings using di-Higgs production at the HL-LHC and how this correlates with the production of detectable GWs, Fig.~\ref{fig:BSMDiHiggsProduction} shows how the power spectrum peak correlates with the size of the di-Higgs production of BSM CP-even Higgs particles. Again, these are the partonic cross-sections for gluon-gluon fusion at fixed $\sqrt{\hat s}=400$ GeV  in the heavy-top quark approximation normalized to the SM value. So it should only be viewed as a diagnostic to compare different scan points. 

Each row in the figure shows a different pairing of BSM Higgs particles. Starting with the non-conformal model we see that for the $ll$ and $HH$ final states the partonic cross-section is typically of the same size as the SM one in the region with a detectable GW signal, making it difficult to observe experimentally unless the BSM Higgses have already been found in single production mode. For the mixed final states $hl$ and $hH$, the partonic cross-section can be substantially larger than the SM one, in particular for $hl$. Turning to the conformal model we see that there are points for all different Higgs pairings that have a substantially larger partonic cross-section compared to the SM, also in the region with detectable GWs.

We emphasize that for a quantitative statement about detectability the full hadronic cross sections, finite-top-mass effects, branching ratios and detector-level final states would have to be calculated. The clear differences between the conformal and non-conformal models provide an additional means to differentiate between the two, as these high production rates were mostly found for the conformal model. Within the present scan, an enhanced partonic BSM Higgs-pair rate therefore appears primarily in the conformal realization, but a full collider study would be required before turning this pattern into an experimental discriminator.

\section{Conclusions}\label{sec::Conclusion}
We have studied the electroweak phase transition and associated stochastic GW signals in the N2HDM, comparing the ordinary non-conformal theory with its classically SI counterpart. After imposing appropriate theoretical constraints as well as collider, flavour and electroweak constraints implemented through \texttt{ScannerS}, \texttt{HiggsBounds} and \texttt{HiggsSignals}, we find viable points in both scenarios with a first-order EWPT and potentially observable sound-wave GW signals, including points in the projected LISA sensitivity region. Within the scanned parameter space the transitions complete promptly, the nucleation and percolation temperatures remaining close ($T_n\simeq T_p$, Fig.~\ref{fig:PTObservables}); pronounced supercooling below the critical temperature ($1-T_p/T_c$ up to $\sim0.8$) is confined to the strongest transitions ($\alpha\gtrsim1$), which are also those for which the sound-wave-only signal-to-noise estimate is least reliable.

The most distinctive result is the correlated pattern between GW observability and the scalar mass spectrum. In the scan performed here, GW-observable conformal and non-conformal points populate different regions of the additional-Higgs mass planes. The charged and CP-odd scalar masses, together with the heavier CP-even spectrum and $t_\beta$, therefore provide a possible way to discriminate between the two realizations if a stochastic background compatible with an electroweak-scale transition was observed. This statement should be understood as a scan-level discriminator, not as a model-independent theorem.

We have also studied fixed-energy partonic Higgs-pair production in the heavy-top approximation as an indicator of complementarity between collider Higgs observables and GW observables. Large deviations in the SM-like Higgs-pair rate tend to occur away from the regions with the largest GW signals, while enhanced BSM Higgs-pair rates arise mainly in the conformal model. A full collider prediction would require finite-top-mass effects, convolution with parton distribution functions, higher-order QCD corrections, scalar branching ratios and detector-level selections, but the present partonic analysis already shows that GW searches and Higgs-pair probes test complementary aspects of the extended scalar potential.

Several theoretical systematic uncertainties should be kept in mind when interpreting these results. The EWPT calculation is based on a one-loop finite-temperature effective potential with a fixed renormalization and resummation prescription, and the extracted transition observables therefore retain a residual gauge and scheme dependence. The GW forecasts use the sound-wave contribution as the baseline source; points with very large $\alpha$ require particular care because the finite lifetime of the acoustic source, turbulence and possible bubble-collision contributions can change the power spectrum. Finally, the Higgs-pair study is performed at partonic level and should be viewed as a complementarity diagnostic rather than as a direct HL-LHC reach projection. These limitations define clear directions for future work, but they do not alter the main qualitative result: conformal and non-conformal N2HDM scenarios can both lead to potentially observable stochastic GW backgrounds, while populating different regions of the scalar-spectrum and Higgs-pair-production parameter space.

\appendix

\section{Conformal Counterterms} \label{sec::AppendixCT}
The conformal CTs are obtained by imposing Eq.~\eqref{eq::ModifiedOS}. It is useful to write the additional conformal contribution in terms of the unit vector along the tree-level flat direction:
\begin{equation}
    n=(n_1,n_2,n_S),\qquad n_i\equiv R_{i1}^{\rm flat}=v_i/v_h,\qquad \sum_i n_i^2=1,\label{eq::flat_direction_vector}
\end{equation}
where the index is the same as in the implementation of the rotation matrix. The modified OS condition is equivalent to adding
\begin{equation}
    \Delta M^2_{ij}=m_{\rm sc}^2 n_i n_j\label{eq::delta_mass_projector}
\end{equation}
to the CP-even scalar mass matrix generated by $V_{CW}+V_{CT}$. Here $m_{\rm sc}\equiv m_{h_1}^{\rm 1-loop}$ denotes the radiatively generated scalon mass in the conformal case. The CT coefficients are then fixed by solving
\begin{align}
    \left.\frac{\partial V_{CW}}{\partial \varphi_i}\right|_0+\left.\frac{\partial V_{CT}}{\partial \varphi_i}\right|_0&=0,\label{eq::ct_tadpole_condition}\\
    \left.\frac{\partial^2 V_{CW}}{\partial \varphi_i\partial\varphi_j}\right|_0+\left.\frac{\partial^2 V_{CT}}{\partial \varphi_i\partial\varphi_j}\right|_0&=\Delta M^2_{ij}.\label{eq::ct_mass_condition}
\end{align}
Compared with the ordinary OS solution of Ref.~\cite{Basler:BSMPT1}, the finite CTs therefore receive shifts whose only role is to generate the radiative scalon mass while preserving the vacuum position. 
In the implementation used for the scan we choose the two remaining free CTs as
\begin{equation}
     \label{eq::ct_free_choice}
    \delta\lambda_4=0,\qquad \delta\lambda_5=\delta\lambda_5^{\rm OS}
\end{equation}
and fix the singlet tadpole CT by requiring that the tadpole part of the CT potential vanish along the VEV direction,
\begin{equation}
    \label{eq::ct_singlet_tadpole}
    \delta T_S=-\frac{v_1}{v_s}\delta T_1-\frac{v_2}{v_s}\delta T_2. 
\end{equation}
The compact projector form in Eqs.~\eqref{eq::delta_mass_projector}--\eqref{eq::ct_mass_condition} provides the convention-independent definition of the modified conformal OS prescription used here.

We use Eqs.~\ref{eq::ct_tadpole_condition}--\ref{eq::ct_mass_condition} as the defining form of the modified conformal on-shell prescription throughout the numerical analysis. This compact form avoids any ambiguity associated with row-column conventions in the CP-even rotation matrix. 
In the implementation, the projector $n_i n_j$ is evaluated in the same real CP-even interaction basis and with the same rotation convention as used for the tree-level mass matrix diagonalization in Sec.~\ref{sec::Model}. 
The expanded component-level CTs are therefore fixed unambiguously by solving Eqs.~\ref{eq::ct_tadpole_condition}--\ref{eq::ct_mass_condition} with the finite-part choices in Eqs.~\ref{eq::ct_free_choice}--\ref{eq::ct_singlet_tadpole}.

For completeness we also give the explicit forms of the counter terms.  Denoting the value obtained in the standard OS scheme in Ref.~\cite{Basler:BSMPT1} as $\delta^{OS}$ and the correction to the scalon mass, from Eq. \ref{eq::ModifiedOS}, as $m_h$ , the conditions are given as follows:\\
Quartic couplings:
\begin{eqnarray*}
    \delta \lambda_1 &=& \delta \lambda_1^{OS} +R_{11}^2\dfrac{m_h^2}{v_1^2}, \\
\delta \lambda_2 &=& \delta \lambda_2^{OS}+R_{21}^2\dfrac{m_h^2}{v_2^2}, \\
\delta \lambda_3 &=& \delta  \lambda_3^{OS}+ R_{11}R_{12}\dfrac{m_h^2}{v_1v_2}, \\
 \delta \lambda_4&=&0, \\
 \delta \lambda_5&=&\delta \lambda_5^{OS}, \\
\delta \lambda_6 &=& \delta  \lambda_6^{OS}+R_{31}^2\dfrac{m_h^2}{v_s^2}, \\
\delta \lambda_7 &=&\delta  \lambda_7^{OS}+R_{11}R_{31}\dfrac{m_h^2}{v_1v_s}, \\
\delta \lambda_8&=&\delta  \lambda_8^{OS}+R_{21}R_{31}\dfrac{m_h^2}{v_2v_s}.
\end{eqnarray*}
Mass-parameters:
\begin{eqnarray*}
\delta m^2_{11} &=&\delta {m^2_{11}}^{OS} -\dfrac{1}{2}m_h^2\left(R_{11}^2+R_{11}R_{21}\dfrac{v_2}{v_1}+R_{11}R_{31}\dfrac{v_s}{v_1}\right), \\
\delta m^2_{22} &=&\delta {m^2_{22}}^{OS}-\dfrac{m_h^2}{2}\left(R_{11}R_{21}\dfrac{v_1}{v_2}+R_{21}^2+R_{21}R_{31}\dfrac{v_s}{v_2}\right), \\
\delta m^2_S &=& \delta {m^2_S}^{OS}-\dfrac{m_h^2}{2}\left(R_{11}R_{31}\dfrac{v_1}{v_S}+R_{21}R_{31}\dfrac{v_2}{v_s}+R_{31}^2\right).
\end{eqnarray*}
Tadpoles:
\begin{eqnarray*}
\delta T_1&=&\delta T_1^{OS}, \\
\delta T_2&=&\delta T_2^{OS},\\
 \delta T_S &=& -\dfrac{v_1}{v_s}\delta T_1- \dfrac{v_2}{v_s}\delta T_2.
\end{eqnarray*}

\section{Details of parameter space scan} \label{sec::AppendixSearch}
In order to explore properties of the conformal and non-conformal N2HDM  we have scanned the respective parameter spaces. For the conformal model we use the parameters 
\begin{equation}
   m_{h_1}, m_{h_2}, m_{h_3},m_{H^\pm},m_A,t_\beta,\alpha_2,\alpha_3,
\end{equation}
whereas for the non-conformal one we also use
\begin{equation}
   \alpha_1,v_s,m_{12}^2 . 
\end{equation}
The bounds used for searching these parameters  are given in Tab. \ref{tab:SearchBoundsC} and \ref{tab:SearchBoundsNC} for the conformal and non-conformal models respectively.  
The scans labeled C-A, C-B, NC-A and NC-B have been used for producing all plots for scenarios A and B respectively, except Fig.~\ref{fig:Theoretical Constraints} which is based on the C-T and NC-T scans.

\begin{table}[ht]
    \centering 
    \caption{The ranges used when scanning the parameter space of the confromal model. A single value indicates that the parameter is fixed to that value. In addition, parameters given with an absolute value means that negative values have also been scanned over.  Scientific notation is used for those parameters that have been scanned with a logarithmic distribution whereas  for the others a uniform distribution is used.  }
    \vspace*{0.2cm}
    \begin{tabular}{|c|c|c|c|}
        \hline
         & \multicolumn{3}{c|}{Scan}\\
         \hline
       Parameter & C-A&C-B&C-T  \\ 
        \hline
        $m_{h_1}$  & 125.1& $[60,200]$ & $[60,400]$ \\ \hline
        $m_{h_2}$  & $[60,200]$&$[60,200]$&$[60,400]$  \\ \hline
        $m_{h_3}$  & $[60,200]$&125.1&$[60,400]$ \\ \hline
        $m_{H^\pm}$  & $[100,450]$ & $[100,450]$&$[100,650]$  \\ \hline
        $m_A$  &$[100,450]$& $[100,450]$&$[100,650]$   \\ \hline
        $t_\beta$ & $[1.5,10]$ & $[1.5,10]$ &  $[10^{-1},10^{1.5}]$   \\ \hline
        $|\alpha_2|$ & $[0,0.5]$ & $[\frac{\pi}{2}-0.5,\frac{\pi}{2}]$ & $[0,\frac{\pi}{2}]$   \\ \hline
        $\alpha_3$ &$[-\frac{\pi}{2},\frac{\pi}{2}]$ & $[-\frac{\pi}{4},\frac{\pi}{4}]$ & $[-\frac{\pi}{2},\frac{\pi}{2}]$  \\ \hline
    \end{tabular}
    \label{tab:SearchBoundsC}
\end{table}

\begin{table}[ht]
    \centering 
    \caption{The ranges used when scanning the parameter space of the non-conformal model. The information is given in the same way as in Tab.~\ref{tab:SearchBoundsC}. }
    \vspace*{0.2cm}
        \begin{tabular}{|c|c|c|c|}
        \hline
         & \multicolumn{3}{c|}{Scan}\\
         \hline
       Parameter & NC-A&NC-B&NC-T \\ 
        \hline
        $m_{h_1}$   & 125.1& $[60,200]$& $[60,400]$\\ \hline
        $m_{h_2}$   & $[60,200]$ & $[60,200]$&$[60,400]$ \\ \hline
        $m_{h_3}$   & $[60,200]$ & 125.1 & $[60,400]$ \\ \hline
        $m_{H^\pm}$  & $[100,450]$ & $[100,450]$& $[100,650]$ \\ \hline
        $m_A$        & $[100,450]$ & $[100,450]$& $[100,650]$ \\ \hline
        $t_\beta$   & $[1.5,10]$ & $[1.5,10]$&$[10^{-1},10^{1.5}]$ \\ \hline
        $|\alpha_2|$ & $[0,0.4]$ & $[\frac{\pi}{2}-0.4,\frac{\pi}{2}]$ & $[0,\frac{\pi}{2}]$ \\ \hline
        $\alpha_3$   & $[-\frac{\pi}{2},\frac{\pi}{2}]$ & $[-\frac{\pi}{4},\frac{\pi}{4}]$ & $[-\frac{\pi}{2},\frac{\pi}{2}]$ \\ \hline
        $\alpha_1$ &$[\beta-0.3,\beta+0.3]$&$[\beta-0.3,\beta+0.3]$&[$-\frac{\pi}{2}$,$\frac{\pi}{2}$]\\ \hline
        $v_s$  & $[24,8\cdot 10^6]$& $[24, 8\cdot 10^6]$ & $[24,8\cdot 10^7]$ \\ \hline
        $m_{12}^2$  & $[5,5\cdot 10^5]$ & $[5,5\cdot 10^5]$ & $[5,5\cdot 10^6]$ \\ \hline
    \end{tabular}
    \label{tab:SearchBoundsNC}
\end{table}

Note that in the conformal case $m_{h_1}$ enters through Eq.~\ref{eq::ModifiedOS} whereas in the non-conformal case it is the tree-level mass. We scan the parameter space as follows:
\begin{enumerate}
     \item Points are found by randomly generating parameter values within the given bounds which then are checked against the theoretical constraints in Sec.~\ref{sec::TheoConstraints}. 
    \item Parameter points that pass all theoretical constraints are then checked against the phenomenological constraints using \texttt{ScannerS}, the details of which are in Sec.~\ref{sec::PhenoConstraints}.
    \item These points are then given to \texttt{BSMPT} which calculates properties of the EWPT and of the primordial GWs as outlined below in section \ref{sec::EWPT}. 
    \item Finally, we calculate the partonic di-Higgs production cross section using Ref.~\cite{Plehn:DiHiggs,Dawson:DiHiggs} as described in section \ref{sec::DiHiggs}.
\end{enumerate}

To convert between the parameters used in the scan and the quartic and quadratic couplings, we use the relations in appendix~\ref{sec::appendixCoupling} which have been derived from the mass matrix. Using these relations as well as calculating $\lambda_6$, $\lambda_7$ and $\lambda_8$ from the tadpole conditions (Eqs.~\ref{eq::tadpole1}-\ref{eq::tadpole3}) gives all tree-level couplings in the potential.

\begin{table}[t]
\centering
\caption{Summary of number of points in the parameter-space scan. ``Theory pass'' denotes points satisfying boundedness from below, perturbative unitarity and the requirement of electroweak symmetry breaking. ``Pheno.~pass'' denotes points also passing the collider, electroweak-precision and flavour constraints. ``Percolates'' denotes points for which a first-order transition is found and completes according to the percolation criterion. Finally, ``SNR$_{\rm LISA}>10$'' denotes the number of points which have SNR$_{\rm LISA}>10$.}
\label{tab:scan_statistics}
\vspace*{0.2cm}
\begin{tabular}{lcccc}
\hline
Scenario  & Theory pass & Pheno.~pass & Percolates & SNR$_{\rm LISA}>10$ \\
\hline
Conformal       & 10 000 000 & 370 000&  67 000   &9  \\
Non-conformal   & 10 000 000 & 130 000 & 61 000  & 39\\
\hline
\end{tabular}
\end{table}

The resulting scan statistics are summarized in Tab.~\ref{tab:scan_statistics}. 
This table is useful for assessing the relative efficiency of the theoretical, phenomenological and finite-temperature requirements in the conformal and non-conformal scans.

\section{Tree-Level Coupling Formulas}\label{sec::appendixCoupling}

Starting from the mass matrix, we find the following formulas to convert from the parameters used in the scan to the couplings in the Higgs potential. It should be noted that $m_{h_i}$ here strictly refers to the tree-level masses and does not use loop-level corrections for the scalon. 

For mass parameters:
\begin{align*}
    m_{11}^2 &= \dfrac{1}{2 v_1} \big( 
- R_{11}^2\, m_{h1}^2\, v_1
- R_{11} R_{12}\, m_{h_1}^2\, v_2
- R_{11} R_{13}\, m_{h_1}^2\, v_s \\
&\quad 
- R_{21}^2\, m_{h_2}^2\, v_1
- R_{21} R_{22}\, m_{h_2}^2\, v_2
- R_{21} R_{23}\, m_{h_2}^2\, v_s \\
&\quad 
- R_{31}^2\, m_{h_3}^2\, v_1
- R_{31} R_{32}\, m_{h_3}^2\, v_2
- R_{31} R_{33}\, m_{h_3}^2\, v_s
+ 2\, m_{12}^2\, v_2 \big), \\[10pt]
m_{22}^2 &= \dfrac{1}{2 v_2} \big( 
- R_{11} R_{12}\, m_{h_1}^2\, v_1
- R_{12}^2\, m_{h_1}^2\, v_2
- R_{12} R_{13}\, m_{h_1}^2\, v_s \\
&\quad 
- R_{21} R_{22}\, m_{h_2}^2\, v_1
- R_{22}^2\, m_{h_2}^2\, v_2
- R_{22} R_{23}\, m_{h_2}^2\, v_s \\
&\quad 
- R_{31} R_{32}\, m_{h_3}^2\, v_1
- R_{32}^2\, m_{h_3}^2\, v_2
- R_{32} R_{33}\, m_{h_3}^2\, v_s
+ 2\, m_{12}^2\, v_1 \big), \\[10pt]
m_S^2 &= \dfrac{1}{2 v_s} \big( 
- R_{11} R_{13}\, m_{h1}^2\, v_1
- R_{12} R_{13}\, m_{h1}^2\, v_2
- R_{13}^2\, m_{h1}^2\, v_s \\
&\quad 
- R_{21} R_{23}\, m_{h_2}^2\, v_1
- R_{22} R_{23}\, m_{h_2}^2\, v_2
- R_{23}^2\, m_{h_2}^2\, v_s \\
&\quad 
- R_{31} R_{33}\, m_{h_3}^2\, v_1
- R_{32} R_{33}\, m_{h_3}^2\, v_2
- R_{33}^2\, m_{h_3}^2\, v_s \big)
\end{align*}
and for quartic couplings, using $\widetilde\mu^2=\dfrac{m_{12}^2}{c_\beta s_\beta}$, to simplify the expressions:
\begin{align*}
\lambda_1 &= \dfrac{1}{v^2 c_\beta^2}
  \Bigl[\, c_{\alpha_1}^2 c_{\alpha_2}^2\, m_{h_1}^2
         + (c_{\alpha_1} c_{\alpha_3} - s_{\alpha_1} s_{\alpha_2} s_{\alpha_3})^2\, m_{h_2}^2 
         + (c_{\alpha_1} s_{\alpha_3} + c_{\alpha_3} s_{\alpha_1} s_{\alpha_2})^2\, m_{h_3}^2 - \widetilde\mu^2\, s_\beta^2 \Bigr], \\[6pt]
\lambda_2 &= \dfrac{1}{v^2 s_\beta^2}
  \Bigl[\, s_{\alpha_1}^2 c_{\alpha_2}^2\, m_{h_1}^2
         + (c_{\alpha_1} s_{\alpha_2} s_{\alpha_3} + s_{\alpha_1} c_{\alpha_3})^2\, m_{h_2}^2 
+ (c_{\alpha_1} c_{\alpha_3} s_{\alpha_2} - s_{\alpha_1} s_{\alpha_3})^2\, m_{h_3}^2
         - \widetilde\mu^2\, c_\beta^2 \Bigr], \\[6pt]
\lambda_3 &= \dfrac{1}{v^2 }
  \Bigl[\, c_{\alpha_1} s_{\alpha_1} c_{\alpha_2}^2\, (m_{h_2}^2 - m_{h_1}^2)
         + (c_{\alpha_1} c_{\alpha_3} - s_{\alpha_1} s_{\alpha_2} s_{\alpha_3})
           (c_{\alpha_1} s_{\alpha_2} s_{\alpha_3} + s_{\alpha_1} c_{\alpha_3}) m_{h_2}^2\notag\\
&\qquad\quad +(c_{\alpha_1} s_{\alpha_3} + c_{\alpha_3} s_{\alpha_1} s_{\alpha_2})
          (c_{\alpha_1} c_{\alpha_3} s_{\alpha_2} - s_{\alpha_1} s_{\alpha_3}) m_{h_3}^2
         - \widetilde\mu^2+2m_{H^\pm}^2 \Bigr], \\[6pt]
\lambda_4&=\dfrac{1}{v^2}\left(\widetilde\mu^2+m_A^2-2m_{H^\pm}^2\right)\\
\lambda_5&=\dfrac{1}{v^2}\left(\widetilde\mu^2-m_A^2\right).
\end{align*}

\bibliographystyle{JHEP}
\bibliography{citation}
\end{document}